\documentclass[12pt]{article}
\pdfoutput=1

\newlength{\abstractwidth}
\usepackage{cancel}
\usepackage{hyperref}
\usepackage{color}
\usepackage{graphicx}
\usepackage{tikz}
\usepackage{cite}

\usepackage{amsmath}
\usepackage{amssymb}
\usepackage{latexsym}
 \usepackage{setspace}
 
 \usepackage{caption}
\usepackage{subcaption}

\numberwithin{equation}{section}

\renewcommand{\thefootnote}{\fnsymbol{footnote}}
\renewcommand{\thanks}[1]{\footnote{#1}}
\newcommand{\starttext}{
\setcounter{footnote}{0}
\renewcommand{\thefootnote}{\arabic{footnote}}}
\newcommand{\bea}{\begin{eqnarray}}
\newcommand{\eea}{\end{eqnarray}}
\newcommand{\be}{\begin{eqnarray}}
\newcommand{\ee}{\end{eqnarray}}

\def\ie{\begin{equation}\begin{aligned}}
\def\fe{\end{aligned}\end{equation}}
\def\half{{\scriptstyle \frac 12}}

\newenvironment{psmallmatrix}
  {\left(\begin{smallmatrix}}
  {\end{smallmatrix}\right)}

\def\ie{\begin{equation}\begin{aligned}}
\def\fe{\end{aligned}\end{equation}}
\def\cG{{\cal A}}

\def\cD{{\cal D}}

\def\cF{{\cal F}}
\def\cG{{\cal G}}

\def\cI{{\cal I}}

\def\cL{{\cal L}}

\def\cN{{\cal N}}
\def\cO{{\cal O}}

\def\cQ{{\cal Q}}

\def\cT{{\cal T}}

\def\Z{{\mathbb Z}}

\def\ZZ{{\mathbb Z}}

\def\nn{\nonumber}

\def\tr{{\rm tr}}
\def\Tr{{\rm Tr}}

\begin{document}

\starttext

\setcounter{footnote}{0}

\begin{flushright}
%\scriptsize 
{\small QMUL-PH-21-39}\\
{\small DCPT-21/17}
\end{flushright}

\vskip 0.3in

\begin{center}
{\bf Exact expressions for $n$-point maximal $U(1)_Y$-violating integrated \\ correlators in $SU(N)$  $\mathcal{N}=4$ SYM}

\vskip 0.2in

{ Daniele Dorigoni$^{(a)}$, Michael B. Green$^{(b)(c)}$  and Congkao Wen$^{(c)}$} 
   
\vskip 0.15in

{\small ($a$) Centre for Particle Theory \& Department of Mathematical Sciences
Durham University,}
\small{ Lower Mountjoy, Stockton Road, Durham DH1 3LE, UK}

\vskip 0.1in

{ \small ($b$) Department of Applied Mathematics and Theoretical Physics }\\
{\small  Wilberforce Road, Cambridge CB3 0WA, UK}

\vskip 0.1in

{\small  ($c$) School of Physics and Astronomy, Queen Mary University of London, }\\ 
{\small  Mile End Road, London, E1 4NS, UK}

\vskip 0.15in

{\tt \small daniele.dorigoni@durham.ac.uk,  M.B.Green@damtp.cam.ac.uk, c.wen@qmul.ac.uk}

\vskip 0.5in

\begin{abstract}
\vskip 0.1in

The exact expressions for integrated maximal $U(1)_Y$ violating (MUV)  $n$-point correlators in $SU(N)$ ${\mathcal N}=4$ supersymmetric Yang--Mills theory are determined. The analysis generalises previous results on the integrated correlator of four superconformal primaries and is based on supersymmetric localisation.   The integrated correlators are functions of $N$ and $\tau=\theta/(2\pi)+4\pi i/g_{_{YM}}^2$, and are  expressed as two-dimensional lattice sums that are modular forms with holomorphic and anti-holomorphic weights $(w,-w)$ where $w=n-4$.  The correlators satisfy Laplace-difference equations  that relate the $SU(N+1)$, $SU(N)$ and $SU(N-1)$ expressions and generalise the equations previously found in the $w=0$ case.  The correlators can be expressed as infinite sums of Eisenstein modular forms of weight $(w,-w)$.  For any fixed value of $N$ the perturbation expansion of this correlator is found to start at order $( g_{_{YM}}^2 N)^w$.  The contributions of Yang--Mills instantons of charge $k>0$ are of the form 
$q^k\, f(g_{_{YM}})$, where $q=e^{2\pi i \tau}$ and  $f(g_{_{YM}})= O(g_{_{YM}}^{-2w})$ when $g_{_{YM}}^2 \ll 1$. Anti-instanton contributions have charge $k<0$ and are of the form $\bar q^{|k|} \, \hat f(g_{_{YM}})$, where  $\hat f(g_{_{YM}}) = O(g_{_{YM}}^{2w})$ when $g_{_{YM}}^2 \ll 1$. Properties of the large-$N$ expansion are in agreement with expectations based on  the low energy expansion of flat-space type IIB superstring amplitudes. We also comment on the identification of  $n$-point free-field MUV correlators with the integrands of $(n-4)$-loop perturbative contributions to the four-point correlator. In particular, we emphasise the important r\^ole of $SL(2, \mathbb{Z})$-covariance in the construction. 

   \end{abstract}

\end{center}

\baselineskip=15pt
\setcounter{footnote}{0}

\newpage

\setcounter{page}{1}
\tableofcontents

\newpage

%%%%%
\section{Overview and outline}
 %%%%%
 
 In  recent work  \cite{Dorigoni:2021bvj, Dorigoni:2021guq}
 we conjectured an exact expression for an integrated four-point correlator of superconformal primaries of the stress tensor multiplet of $\cN=4$ $SU(N)$ supersymmetric Yang--Mills theory, that  is given by a two-dimensional lattice sum and manifests the $SL(2, \mathbb{Z})$ modular symmetry of the theory.  In this paper we will extend these results to $n$-point correlation functions that violate $U(1)_Y$ charge conservation maximally.
 
 \subsection{Overview}
 
The standard correlators of operators in supersymmetric conformal field theory are position dependent and therefore in general break supersymmetry.  However,  integrating over the positions of the operators in a correlator with suitable measure leads to a supersymmetric integrated correlator. The form of certain integrated  correlators in $\cN=4$ supersymmetric $SU(N)$ Yang--Mills (SYM) theory can be determined by supersymmetric localisation in the manner described in  \cite{Binder:2019jwn} and briefly reviewed in appendix~\ref{review}.  These are obtained by exploiting the fact that $\cN=4$ SYM theory is a limit of $\cN=2^*$ SYM theory in which the hypermultiplet mass vanishes.  The particular integrated correlator considered in the large-$N$ expansion in  \cite{Binder:2019jwn} is the correlator of four superconformal primaries, $\langle \cO_2(x_1, Y_1)\cdots \cO_2(x_4, Y_4)\rangle$, integrated over their positions, $x_i$, with a particular measure.\footnote{Here $Y_i$ is a $SO(6)$ null vector, encoding the R-symmetry information of $\cN=4$ SYM.  This dependence in the correlator can be factored out and is described in appendix~\ref{review}.}   
This is given  by taking four derivatives of the logarithm of the partition function of $\cN=2^*$ $SU(N)$ SYM on $S^4$,  
  \bea \label{integratedF1}
\cG_{N}(\tau,\bar\tau) = \left.  {1\over 4} \, { \Delta_{\tau}\partial_m^2 \log Z_{N}}(m, \tau, \bar{\tau})    \right |_{m=0}   \, ,
\label{firstmeasure}
\eea
where $\Delta_{\tau}=4 \tau_2^2\partial_\tau\partial_{\bar\tau}$ is the hyperbolic Laplacian. The partition function of $\cN=2^*$  SYM, 
$Z_N(m,\tau,\bar\tau)$, is precisely determined by supersymmetric localisation \cite{Pestun:2007rz}.   The parameter $m$ is a mass parameter and in the limit in which it vanishes ($m=0)$, the hypermultiplet mass vanishes and $\cN=2$ supersymmetry is extended to $\cN=4$.  

Our notation follows usual conventions where the complex Yang-Mills coupling constant is defined by
\begin{equation}
\tau=\tau_1 + i \tau_2 := \frac{\theta}{2\pi}+i  \frac{4\pi^2}{g_{_{YM}}^2}\,,
\end{equation}
with $\theta$ the topological theta angle and $g_{_{YM}}$ the Yang-Mills coupling constant.

The large-$N$  
't Hooft expansion (in which the 't Hooft coupling $\lambda=g_{_{YM}}^2 N$ is fixed) of $\cG_N(\tau,\bar\tau)$ was considered in some detail  in \cite{Binder:2019jwn, Chester:2019pvm}.  The large-$N$ expansion with fixed $g_{_{YM}}^2$ was considered in \cite{Chester:2019jas}, where the instanton contributions to the correlator play an essential r\^ole in implementing Montonen--Olive $SL(2,\Z)$ duality \cite{Montonen:1977sn, Witten:1978mh, Osborn:1979tq}.
The considerations in  \cite{Dorigoni:2021bvj, Dorigoni:2021guq} led to a reformulation of this correlator as a two-dimensional lattice sum,  which makes  the modular properties of $\cG_N(\tau,\bar\tau)$ manifest  for all values of $N$ and greatly simplifies and extends the analysis of the large-$N$ expansion. These properties of the four-point correlator are also briefly summarised in appendix~\ref{review}.

A second example of an integrated correlator presented in \cite{Chester:2020dja} is obtained from four derivatives with respect to the masses, $ \partial_m^4 \log Z_ N (m, \tau, \bar{\tau})  |_{m=0} $ of $\cN=2^*$  SYM partition function.  and is again an integral of $\langle \cO_2(x_1, Y_1)\cdots \cO_2(x_4, Y_4)\rangle$ over $x_i$, but with a different measure.  Its large-$N$ expansion in the 't Hooft and fixed $g_{_{YM}}^2$ limits were discussed in \cite{Chester:2020dja} and \cite{Chester:2020vyz}, respectively.  The exact results of these integrated correlators have been used to determine scattering amplitudes of type IIB superstring theory in $AdS_5 \times S^5$, after taking flat-space limit, which match precisely with known results \cite{Green:1997tv, Green:1997as, Green:1998by, Green:1999pu, Green:2005ba}.  

\vskip 0.3cm
{\bf $U(1)_Y$-violating correlators}
\vskip 0.2cm

Our aim here is to extend the preceding considerations to a class of $n$-point correlators of operators in the stress tensor supermultiplet that are modular forms with non-zero modular weights $(w,-w)$, so they transform under $SL(2,\Z)$  by a $U(1)_Y$ transformation, where the $U(1)_Y$ charge is given by $q_{_{U}}= 2w$ (see appendix \ref{app:Eisendef} for a brief summary of some relevant $SL(2,\Z)$ properties).
Here $U(1)_Y$, which was called the `bonus $U(1)_Y$' in \cite{Intriligator:1998ig}, is the holographic image of the $U(1)$ R-symmetry in type IIB supergravity and breaks to $\mathbb{Z}_4$ when stringy corrections are turned on. 

The $U(1)_Y$ charge of a correlation function of operators in the stress tensor supermultiplet is the sum of the charges of the individual operators in the correlator. Any of these super-descendent operators has the form $\delta^n \bar\delta^{\hat n} \cO_2$, where $\delta$ is a chiral supersymmetry transformation carrying  $U(1)_Y$ charge $+1/2$  and $\bar \delta$  is an anti-chiral supersymmetry transformation with $U(1)_Y$ charge $-1/2$.  
  Since the superconformal primary, $\cO_2(x,Y)$ has zero $U(1)_Y$ charge these descendants $\delta^n \bar\delta^{\hat n} \cO_2$ possess a charge equal to $(n-\hat n)/2$.     Furthermore, the stress tensor supermultiplet is ultra-short so that $n+\bar  n \le 4$.

Super-descendent operators of particular significance in the following are the chiral and anti-chiral Lagrangian operators, $\cO_\tau = \delta^4 \cO_2$ and $\bar \cO_{\bar \tau} = \bar \delta^4 \cO_2$, which  carry $U(1)_Y$ charge $+2$ and $-2$, respectively.
The $\cN=4$ SYM Lagrangian can be expressed as the sum of two complex conjugate parts 
\begin{align}
\cL=&-   \frac{i}{2 \tau_2}  \left( \tau \cO_\tau-  \bar \tau  \bar\cO_{\bar \tau}\right)\,, 
\label{loperdef}
\end{align}
where the chiral and anti-chiral Lagrangians are defined by  
\begin{align}
 \label{otaudef} 
 \cO_\tau = \frac{ \tau_2}{4\pi}  \,  \tr \left( -{1\over 2}  F_{\alpha\beta} F^{\alpha\beta} +\dots \right)\,,\qquad\qquad
\bar\cO_{\bar\tau} = \frac{ \tau_2}{4\pi} \,\tr \left( - {1\over 2} \bar F_{{\dot \alpha}{\dot \beta}} \bar  F^{{\dot \alpha}{\dot \beta}} 
+\dots  \right)\, ,
\end{align}
where $F_{\alpha\beta} $, $\bar F_{{\dot \alpha}{\dot \beta}}$ are the self-dual and anti self-dual Yang--Mills field strengths and ``$\dots$" indicates terms involving fermion and scalar fields in the Yang--Mills supermultiplet.

 The pattern of $U(1)_Y$ symmetry breaking in the  low energy expansion of type IIB superstring amplitudes was discussed in \cite{Green:1999qt}  and in the large-$N$ expansion of  $\cN=4$ SYM in \cite{Intriligator:1998ig} (see also \cite{Eden:1999gh}).
In either case, this symmetry is broken so that in general the $U(1)_Y$ charge is violated by the $n$-point correlator with $n>4$.  The  magnitude of the modular weight of any such correlator has an upper bound given by $|w|\le n-4$.
This means that the maximum $U(1)_Y$ charge violation is given by 
 \bea
|q_U| = 2|w|= 2n-8\,.
\label{maxw}
\eea

\vskip 0.3cm
{\bf Maximal $U(1)_Y$-violating correlators}
\vskip 0.2cm

Maximal $U(1)_Y$-violating (MUV) correlators are $n$-point correlation functions with maximal $U(1)_Y$ charge, i.e. $q_U=2n-8$, that have modular weights $(w,-w)$, where $w=n-4$.  General features of  such correlators were considered in \cite{Green:2020eyj} where particular emphasis was placed on those terms in the large-$N$ expansion that correspond to the BPS protected terms in the low energy expansion of the holographic dual type IIB string theory studied in \cite{Green:2019rhz}.  
A characteristic feature of such MUV string amplitudes is that they do not possess massless poles in any channel.\footnote{``Next-to-MUV''  (NMUV) $n$-point correlators were also defined in  \cite{Green:2019rhz}.  These are dual to type IIB string theory amplitudes that have a massless pole in one channel with a residue that is the product of a $(n-1)$-point MUV amplitude and a three-point supergravity vertex.  Furthermore, ``Next-to-next-to-MUV'' (NNMUV) $n$-point correlators were defined to be correlators that  are dual to string amplitudes in which a massless pole either has a residue proportional to the product of a $n_1$-point and a $n_2$-point MUV amplitude (with $n_1+n_2=n-2$ and $n_1,n_2\ge 4$) or into the product of a three-point supergravity amplitude and a NMUV $(n-1)$-point amplitude.} 
This extended the analysis of the large-$N$ expansion of the correlators in \cite{Binder:2019jwn,Chester:2019pvm,Chester:2019jas,Chester:2020dja, Chester:2020vyz}   to $n$-point MUV correlators.

 There is a convenient harmonic superspace description which packages MUV correlators  together \cite{Howe:1999hz, Eden:2011we}.  
In this approach the operators in the stress-tensor super-multiplet, $\cT(\Psi)$, are functions of the superspace variables $\Psi=(x,y, \rho,\bar\rho)$, where the $y$-dependence determines the dependence on $Y$ (the R-symmetry $SU(4)$)\footnote{The coordinate $y^a_{a'}$ is related to the $SO(6)$ null vector $Y_I$ by $Y_I =  (\Sigma_I)^{AB} \epsilon_{ab} g^a_A g^b_B / \sqrt{2}$, 
where $g_A^b = (\delta^b_a, y^b_{a'})$,
which implies 
$(Y_i)_I  \, (Y_j)_I=  (y_i - y_j)^2 $.}       and $\rho$, $\bar \rho$ are Grassmann coordinates in the $\bf{(2,1)}_1$ and  $\bf{(1,2)}_{-1}$ representations of $SU(2) \times SU(2)'\times U(1)\subset SU(4)$.\footnote{The $U(1)$ factor is a subgroup of the $SU(4)$ R-symmetry and should not be confused with the $U(1)_Y$ bonus symmetry, which  is an automorphism of $PSU(2,2|4)$, and which is broken to $\Z_4$.}    The $n$-point correlators of interest to us are the  coefficients in the expansion of the correlator of $n$ $\cT$'s  in powers of $\rho_i$ and $\bar \rho_i$,\footnote{We have introduced a small change in the notation used in \cite{Green:2020eyj}.}
\bea
 \langle  \mathcal{T}(\Psi_1)  \mathcal{T}(\Psi_2) \cdots  \mathcal{T}(\Psi_n) \rangle
 = \!\!\! \sum_{\underset{  |\sum_{r=1}^n (k_r -  \bar \ell_r)| \,    \leq\, 4n-16 } { \{ k_r,  \bar \ell_r\}=0}}^4 \!\!\! \widehat G^{(w)}_N (j_1, j_2, \dots, j_n) \,  \rho_1^{k_1} \bar\rho_1^{ \bar \ell_1}\cdots \rho_n^{k_n} \bar\rho_n^{ \bar \ell_n}  \, ,
\label{stresscorr}
\eea
where the variables denoted by each label are  $j_i = (x_i,y_i,k_i,  \bar \ell_i)$ and the superscript $(w)$ indicates the modular weight of the correlator, which equals to $w=\sum_{r=1}^n (k_r -  \bar \ell_r)/4$.
The fact that the stress-tensor multiplet is ultra-short implies that the sums are subject to the restrictions 
\bea
k_r+  \bar \ell_r \le 4\,,  \qquad  \quad (k_r,\ell_r) \ne (1,3) \ {\rm or} \ (3,1)\,, \qquad\quad  {\rm where}\    
1\leq r \leq n,
\label{klcon}
\eea
and furthermore, as explained in \cite{Eden:2011we}, supersymmetry and superconformal symmetry imply that
$\bigg|\sum_{r=1}^n k_r - \sum_{r=1}^n  \bar \ell_r\bigg|   \le 4n-16$. 
 The correlator $ \widehat G^{(w)}_N (j_1, j_2, \dots, j_n)$ is a correlator of super-descendants of the form 
 \bea
 \widehat G^{(w)}_N (j_1, j_2, \dots, j_n)= \langle \cO_{k_1,  \bar \ell_1}(x_1,Y_1)  \cO_{k_2,  \bar \ell_2}(x_2,Y_2)\cdots  \cO_{k_n,  \bar \ell_n}(x_n,Y_n)  \rangle\,,
 \label{correx}
 \eea
 where $(k_r,  \bar \ell_r)$ label the components of the stress tensor super-multiplet (for example,  $\cO_2 \equiv \cO_{0,0}$, $\cO_\tau\equiv \cO_{4,0}$ and $\bar \cO_{\bar \tau} \equiv \cO_{0, 4}$). For the MUV correlators, we have $\sum_{r=1}^n k_r - \sum_{r=1}^n  \bar \ell_r  = 4n-16$, or equivalently $w=n-4$. 
   
 For much of the following we will restrict our considerations to MUV correlators of chiral operators, which have  the form $\delta^n \cO_2$ (with $n\le 4$), in which case $ \bar \ell_r=0$ in \eqref{correx}.
One example of such a correlator, which is particularly relevant in the following discussion, is
\bea
 \label{relevant}
\langle   \cO_2(x_1, Y_1) \cdots  \cO_2(x_4, Y_4)\, \cO_\tau(x_5)\cdots \cO_\tau(x_{4+m})\rangle \,,
\label{mcorr}
\eea
which is the four-$\cO_2$ correlator with $m$  insertions of $\cO_\tau$.  Each insertion increases the modular weight $w$ by $1$, so that the total weight of the correlator is $w=m$.   
This correlator  is related by superconformal symmetry to other MUV correlators with the same modular weight (or equivalently the same number of operators). For example, when $m=12$  \eqref{mcorr} is  related to the product of sixteen fermionic operators of the form  $\langle \Lambda(x_1,Y_1) \Lambda(x_2,Y_2)\cdots \Lambda(x_{16},Y_{16}) \rangle$, where the descendent $\Lambda\sim \delta^3 \cO_2=\cO_{3,0}$ has $U(1)_Y$ charge $3/2$ so this correlator has $U(1)_Y$ charge $24$ ($w=12$). This is the holographic dual of the sixteen-dilatino interaction in type IIB superstring theory.

An important property of chiral MUV correlators is that they can be written in the form
\bea
 \widehat G^{(n-4)}_N (j_1, j_2, \dots, j_n)= \cI_n^{k_1,\dots,k_n}(x_1,\dots, x_n; y_1,\dots, y_n)\, G_N^{(n-4)}(x_1,\dots, x_n; \tau,\bar\tau)\,,
\label{ienextra}
\eea
and so the dependence on the operator content of the correlator  (including the $SU(4)$ quantum numbers) is contained in the pre-factor factor $ \cI_n^{k_1,\dots,k_n}(x_1,\dots, x_n; y_1,\dots, y_n)$, which is fixed by the symmetries and independent of the coupling.   This is the generalisation of the  factor of $\cI_4(U,V;Y)$ in the case of the four-point function in \eqref{corrdefs}. The remaining factor, $G^{(n-4)}_N(x_1,\dots, x_n; \tau,\bar\tau)$ is the ``reduced correlation function''  that has the same form for any  MUV $n$-point correlator, and is the analogue of $\cT_N(U,V)$ in \eqref{corrdefs}. The explicit expression for  $ \cI_n^{k_1,\dots,k_n}(x_1,\dots, x_n; y_1,\dots, y_n)$ was determined in \cite{Eden:2011we} and is reproduced in section 2 of  \cite{Green:2020eyj} (where references to the original observations can be found).    The fact that MUV correlators of a given modular weight are explicitly related by supersymmetry is the analogue of the property of MUV superamplitudes in type IIB string theory.  There, the $n$-point amplitudes possess an overall prefactor of $\delta^{16}(\sum_{i=1}^n Q_i)$, where $Q_i$ is the sixteen-component supercharge acting on the $i^{th}$ particle.\footnote{See \cite{Abl:2020dbx} for a recent application of this observation in the study of low-energy expansion of superamplitudes in type IIB superstring theory in $AdS_5 \times S^5$. }  The challenge is to determine properties of the reduced correlation function, $G^{(n-4)}_N(x_1,\dots, x_n; \tau,\bar\tau)$.

General properties of  $G^{(n-4)}_N(x_1,\dots, x_n; \tau,\bar\tau)$ and its large-$N$ expansion at finite coupling $\tau$ were studied in detail in  \cite{Green:2020eyj}.  A key result is obtained by applying the $SL(2, \mathbb{Z})$ covariant derivative, $\cD_w$, to a correlator. This acts on  the  factor of $e^{\int \!d^4x \,\cL(x)}$ in the definition of the expectation value \eqref{corredef}, thereby inserting an integrated chiral lagrangian,  $\int \!d^4 x \,\cO_\tau (x)$. Care must be taken to include the contributions of the integrated contact terms arising from this insertion, which have the form  $\int \!d^4 x\, \cO_\tau (x)  \cO_{w_r}(x_r) \sim -(1+w_r) \, \cO_{w_r}(x_r)$, for each operator $\cO_{w_r}(x_r)$ in the correlator with modular weight $(w_r,-w_r)$ (as discussed in  \cite{Basu:2004dm, Basu:2004nt, Green:2020eyj}). The derivative also acts on the factor of $\tau_2$ in the normalisation of each of the  operators in the correlator.   
 The net result is the recursion relation
  \bea
   \label{eq:soft}
 \cD_w\, G_N^{(n-4)}(x_1, \dots , x_{n}; \tau,\bar\tau)=  {1\over 2}   \int {d^4 x_{n+1} } \,G_N^{(n-3)} (x_1, \dots , x_n, x_{n+1}; \tau,\bar\tau) \, ,
 \eea
 which expresses the content of a soft dilaton condition in the dual holographic superstring theory.  Here $w=n-4$ and the covariant derivative $ \cD_w$ is defined as
 \ie
 \cD_w =  i \Big(\tau_2 \frac{\partial}{\partial \tau}  - i \frac{w}{  2}\Big) \, ,
 \fe
 which acts on a modular form of weights $(w, \hat w)$ and changes it to be a modular form with weights $(w+1, \hat w-1)$.  Thus,  the application of $\cD_w$ to a correlator of weight $(w,-w)$ results in the insertion of $\int \!dx \,\cO_\tau(x)$, which shifts $w$ to $w+1$. Detailed properties of $\cD_w$ are discussed in appendix \ref{app:Eisendef}.

We are here interested in the integrated MUV correlators that generalise $\cG_N(\tau,\bar\tau) \equiv \cG_N^{(0)}(\tau,\bar\tau)$ by the insertion of multiple factors of the integrated chiral Lagrangian, $\int \!dx\, \cO_\tau(x)$.  Such insertions are obtained by applying multiple covariant derivatives $\cD_w$ to $\cG_N(\tau,\bar\tau)$.   The resulting expression is a $(w,-w)$ modular form given by
 \begin{equation}
\cG_N^{(w)}(\tau,\bar\tau) = 2^w\, \cD_{w-1} \cD_{w-2} \cdots \cD_{0}\, \cG_N(\tau,\bar\tau) \, ,
\label{muvres}
\end{equation}
which is in accord with the soft dilaton properties of the holographically conjugate type IIB amplitudes \cite{DiVecchia:2015jaq, Green:2019rhz}, as argued in \cite{Green:2020eyj}.  

The leading terms in the large-$N$ expansion of the MUV correlators that were studied in \cite{Green:2020eyj} have a holographic correspondence with the BPS protected terms in the low energy expansion of the MUV amplitudes in type IIB superstring  studied in \cite{Green:2019rhz}.  These are terms with dimension up to 14, i.e. up to the dimension of the $d^6R^4$ interaction in the $w=0$ sector.
In the following we will generalise the lattice expression \eqref{gsun} to the expression that describes MUV correlators  and determine their behaviour in various limits.
 
\vskip 0.3cm

\subsection{Outline of paper}
\label{outline}
 
In section \ref{sec:exact} we will consider features of integrated $n$-point MUV correlators for general values of $N$, which extend the results of the $n=4$ case.  
For example, in section~\ref{sec:lap} we will show that the correlator $\cG_N^{(w)}(\tau,\bar\tau)$ ($w=n-4$)  defined in \eqref{muvres} satisfies a Laplace-difference equation, which follows directly from the equation satisfied by the four-point correlator $\cG_N(\tau,\bar\tau)$.  We will demonstrate in section~\ref{sec:YMpert}  that a weight-$w$ MUV  correlator can be expressed as a two dimensional lattice sum, extending the analysis of the $w=0$ case given in  \cite{Dorigoni:2021bvj, Dorigoni:2021guq}.  This lattice sum can also be expressed as an infinite sum of Eisenstein modular forms (which are defined and summarised in appendix~\ref{app:Eisendef}).  The structure of the perturbative expansion of  $\cG_N^{(w)}(\tau,\bar\tau)$  in powers of $\lambda= g_{_{YM}}^2 N$ is determined to any desired order for any value of $N$.  As in the $w=0$ case the perturbation series contains non-planar contributions, which start at $(4-w)$-loop order when $w < 4$ and at free theory if $w \geq 4$. 
 
Instanton and anti-instanton contributions are extracted from the exact expression for the correlator in section~\ref{sec:instanton}. Unlike in the  $w=0$ case, when $w>0$ the systematics of the perturbation expansion around an instanton  is different from that around an anti-instanton.    This will be seen to be in accord with semi-classical arguments concerning the fermionic zero modes contained in the profile of the operators in the correlator in an instanton or anti-instanton background.

The large-$N$ expansion of $\cG_N^{(w)}(\tau,\bar\tau)$ is discussed in section~\ref{largeN} where we will determine both the fixed $\lambda$ and fixed $g_{_{YM}}^2$ expansions and demonstrate the similarities and differences from the $w=0$ case. 
At small $\lambda$ we find a convergent perturbative expansion for $|\lambda|<\pi^2$, while for $\lambda\gg1$ perturbation theory produces an asymptotic, factorially growing, divergent series.  This strong coupling series is not Borel summable and its non-perturbative completion, which behaves as ${O}(\lambda^{w/2} e^{-2\sqrt{\lambda}} )$, is determined using resurgence techniques. 

 In  section~\ref{reloop} we will briefly discuss the insertion of $\int \!d^4x \,\cO_\tau(x)$ in the non-integrated correlator $\langle \cO_2(x_1, Y_1)\cdots \cO_2(x_4, Y_4)\rangle$, and its application for constructing perturbative loop integrands \cite{Eden:2011we, Eden:2012tu}. We will argue that it is important to use the covariant derivatives (rather than the ordinary derivatives with respect to $g_{_{YM}}$)
in this procedure of determining perturbative loop integrands of the $4$-point correlator.

We end with a conclusion and discuss some future directions in section~\ref{discuss}.

\section{Exact properties of MUV correlators}
\label{sec:exact}

We will now consider properties of the MUV correlators that are obtained from  \eqref{muvres} using the exact expression for $\cG_N^{(0)}(\tau,\bar\tau) \equiv \cG_N(\tau,\bar\tau)$ in \eqref{gsun}.

\subsection{The Laplace-difference equation}
\label{sec:lap}

It is straightforward to determine the Laplace equation satisfied by  $\cG_N^{(w)}(\tau,\bar\tau)$ given the Laplace difference equation for the $w=0$ case in \eqref{lapdiff}, which was derived in \cite{Dorigoni:2021bvj, Dorigoni:2021guq} and we rewrite here for convenience:
\begin{equation}
(\Delta_\tau-2)\mathcal{G}_N = N^2 ( \mathcal{G}_{N+1}-2 \mathcal{G}_{N}+ \mathcal{G}_{N-1}) -N( \mathcal{G}_{N+1}- \mathcal{G}_{N-1})\,,
\label{lapdifftwo}
\end{equation}
where $\cG_N^{(0)}(\tau,\bar\tau) \equiv \cG_N(\tau,\bar\tau)$.
As described in appendix \ref{app:Eisendef}, the hyperbolic Laplacian $\Delta_{\tau}$ acting on $\cG_N(\tau,\bar\tau)$ can be identified with the $SL(2)$ Casimir operator $\Omega_{(0,0)}$, defined in \eqref{cassimirdef}, when restricted to the space of modular functions, i.e. modular forms $M_{(w,\hat{w})}$ with holomorphic and anti-holomorphic weights $(w,\hat{w})=(0,0)$ .

From equation \eqref{muvres}, we know that $\cG_N^{(w)}$ is obtained by repeated applications of the covariant derivative  to $\cG_N^{(0)}$.
Furthermore, the covariant derivative changes the modular weights according to $\cD_w: M_{(w,\hat{w})}\mapsto M_{(w+1,\hat{w}-1)}$, and since the Casimir operator $\Omega$ commutes with $\cD_w$,  using \eqref{cassimirdef}  it follows that 
\begin{align}
\Omega_{w,-w}  \, \cG_N^{(w)}&\notag= \Omega_{w,-w}\Big[2^w  \mathcal{D}_{w-1} \mathcal{D}_{w-2} \cdots \mathcal{D}_{0} \,\mathcal{G}_N\Big]  = 2^w \mathcal{D}_{w-1} \mathcal{D}_{w-2} \cdots \mathcal{D}_{0}\, \Big[\Omega_{0,0}\, \mathcal{G}_N \Big]\\
& = N^2  ( \cG_{N+1}^{(w)}-2   \cG_N^{(w)}+  \cG_{N-1}^{(w)} )-N(\cG_{N+1}^{(w)}-   \cG_{N-1}^{(w)} )+2  \cG_N^{(w)}\,,
\label{neweq}
\end{align}
where $\Omega_{w,-w} $ denotes the restriction of the Casimir operator to the vector space of modular forms $M_{(w,-w)}$ with weights $(w,-w)$.  The second line follows from the Laplace-difference equation \eqref{lapdifftwo} satisfied by $\mathcal{G}_N(\tau, \bar \tau)$, and the fact that $\Omega_{0,0} = \Delta_{\tau}$.

Given the explicit forms of $\Omega_{w,-w}$ in \eqref{eq:Lapwmw1} and \eqref{eq:Lapwmw2} \eqref{neweq} can  be expressed in either of two ways:
\begin{equation}
\Big(4 \mathcal{D}_{w-1}\bar{\mathcal{D}}_{-w} +[w(w-1)-2] \Big)   \cG_N^{(w)}= N^2 (   \cG_{N+1}^{(w)}-2  \cG_N^{(w)} +   \cG_{N-1}^{(w)})-N(   \cG_{N+1}^{(w)}-   \cG_{N-1}^{(w)})\,,
\label{lapdiffw1}
\end{equation}
or equivalently
\begin{equation}
\Big(4 \bar{\mathcal{D}}_{-w-1}\mathcal{D}_{w} +[w(w+1)-2] \Big)  \cG_N^{(w)}  = N^2 (   \cG_{N+1}^{(w)}-2  \cG_N^{(w)} +   \cG_{N-1}^{(w)})-N(   \cG_{N+1}^{(w)}-   \cG_{N-1}^{(w)})\,.
\label{lapdiffw2}
\end{equation}

Just as in the $w=0$ case described in \cite{Dorigoni:2021bvj, Dorigoni:2021guq}, this equation determines  $\cG_{N}^{(w)}(\tau,\bar\tau)$  for $N>2$ in terms of the $N=2$ MUV integrated correlator, $\cG_{2}^{(w)}(\tau,\bar\tau)$.
It is easy to see that in the perturbative sector, where there is no dependence on $\tau_1$, the operator on the left-hand side of \eqref{lapdiffw1}  reduces to 
\bea
4 \cD_{w-1}\, \bar{\mathcal{D}}_{-w} +[w(w-1)-2]    \rightarrow \tau_2^2 \partial^2_{\tau_2} -2
\label{pertred}
\eea
which is identical to the differential operator of the $w=0$ case.  The same is true for the operator on the left-hand side of \eqref{lapdiffw2}.  In other words, the perturbative part of the Laplace-difference equation is not sensitive to the value of $w$.  This does not mean that  the perturbative part of $\cG_{N}^{(w)}(\tau,\bar\tau)$ is identical to that of $\cG_{N}(\tau,\bar\tau)$, since the  inputs from the $SU(2)$ cases are different ($\cG_{2}^{(w)}(\tau,\bar\tau)$ is different from $\cG_{2}(\tau,\bar\tau)$ for $w>0$). In the following section, we will explicitly discuss the perturbative expansion of $\cG_{N}^{(w)}(\tau,\bar\tau)$. 

\subsection{Yang--Mills perturbation theory}
\label{sec:YMpert}

The expression for $\cG_N^{(w)}(\tau,\bar\tau)$  can be obtained by substituting the expression for $\cG_N(\tau,\bar\tau)$   in terms of non-holomorphic Eisenstein series \eqref{guborelN-3} into \eqref{muvres}, giving
\ie
\mathcal{G}_N^{(w)}(\tau,\bar\tau) &=\frac{N(N-1) }{8}\delta_{w,0}+ \frac{1}{2}\sum_{s=2}^\infty c_s^{(N)} 2^w\, \mathcal{D}_{w-1} \mathcal{D}_{w-2} \cdots \mathcal{D}_{0}\, E(s;\tau,\bar\tau) \\
&= \frac{N(N-1)}{8}\delta_{w,0}+ \frac{1}{2}\sum_{s=2}^\infty c_s^{(N)}\, \frac{1}{\tau_2^{w}} \nabla^w \,E(s;\tau,\bar\tau) \, ,
\label{modcorr}
\fe
where $\delta_{w,0}$ denotes the Kronecker delta and $\nabla = 2 i \tau_2^2 \partial_\tau$ is the Cauchy-Riemann derivative discussed in appendix~\ref{app:Eisendef}. Using the relation $\tau_2^{-w} \,\nabla^w \,E(s;\tau,\bar\tau) =  (s)_w \,E^{(w)} (s;\tau,\bar\tau)$, where $E^{(w)} (s;\tau,\bar\tau)$ is the Eisenstein modular forms that is discussed in appendix~\ref{app:Eisendef}  (and defined by \eqref{zwdef}), one can further express the integrated correlator as
\ie\label{eq:GNWEis}
\mathcal{G}_N^{(w)}(\tau,\bar\tau)  =\frac{N(N-1) }{8}\delta_{w,0}+  \frac{1}{2}\sum_{s=2}^\infty c_s^{(N)}\,  (s)_w \,E^{(w)} (s;\tau,\bar\tau)\, .
\fe
Now, using the Lattice sum expression of $E^{(w)} (s;\tau,\bar\tau)$ given in \eqref{intmodeins} together with \eqref{bexpand}, we find that the weight-$w$ integrated correlator can be further expressed as
\begin{equation}
\cG_N^{(w)} (\tau,\bar\tau)= \sum_{(m,n)\in\mathbb{Z}^2} \frac{d^{2w}}{d\alpha^{2w}}\Big[ \int_0^\infty \exp\Big( - \frac{\pi t | m +n\tau|^2}{\tau_2}+  \alpha \frac{\sqrt{\pi}(m+n\bar{\tau})}{\sqrt{\tau_2}}\Big)  t^w B_N(t) dt\Big]_{\alpha=0}\,.
\label{lattsum}
\end{equation}
This lattice sum representation  is a well-defined analytic modular form for all values of $\tau$ with $\tau_2>0$.
If the rational function $B_N(t)$ is expanded around the origin, as in \eqref{bexpand}, and if we restrict our attention to a single monomial term of the form $t^{s-1}$ we obtain the same integrand as that of $E^{(w)}(s;\tau,\bar\tau)$ in \eqref{intmodeins}. In other words, \eqref{lattsum} can formally be expanded as an infinite sum of $E^{(w)}(s;\tau,\bar\tau)$ modular forms with rational coefficients. 

\subsubsection{The relationship between weak and strong coupling}

The perturbative terms in the small-$g_{_{YM}}^2$ limit can be extracted by proceeding as in the $w=0$ case considered in \cite{Dorigoni:2021bvj, Dorigoni:2021guq}.  Recall that  the perturbative terms come from the zero mode of non-holomorphic Eisenstein series as in \eqref{guborelN-3}, which is given by the sum of two pieces shown in  \eqref{eq:SUN0}.   The sum of the $\tau_2^{1-s}= (g_{_{YM}}^2/4\pi)^{s-1}$ terms is denoted $\cG_{N,0}^{(i)}(\tau_2)$; the other piece is $\cG^{(ii)}_{N,0}(\tau_2)$, which is given by the sum of the  $\tau_2^{s}= (g_{_{YM}}^2/4\pi)^{-s}$ terms. After Borel summation we saw that $\cG^{(ii)}_{N,0}(\tau_2)= \cG^{(i)}_{N,0}(\tau_2)$ and both parts of the zero mode sum contribute equally.

We will now see how this extends to  MUV correlators starting from the expression \eqref{lattsum} and following closely the procedure used to analyse the modes of $E^{(w)}(s;\tau,\bar\tau)$ in appendix~\ref{app:Eisendef}.  When  $w>0$ the term $(m,n)=(0,0)$ is absent since it is killed by the $\alpha$ derivative in \eqref{lattsum}. The Fourier expansion of  \eqref{lattsum} is again obtained by performing a Poisson resummation in $m$ (and later changing $n\to-n$), resulting in
\bea
&&\cG_N^{(w)} (\tau, \bar \tau) \! = \!\! \sum_{(\hat{m},n)\in\mathbb{Z}^2}  \sqrt{\tau_2} e^{ 2\pi i \hat{m} n \tau_1}  \\
&& \qquad \frac{d^{2w}}{d\alpha^{2w}}\Big[ \int_0^\infty \exp\Big( -\frac{(2n \sqrt{\pi \tau_2} t+i\alpha)^2}{4t}-\frac{(2\hat{m}\sqrt{\pi \tau_2}-i\alpha)^2}{4t}-\frac{\alpha^2}{4t} \Big)  t^{w-1/2} B_N(t) dt\Big]_{\alpha=0}\, , \nonumber
\label{eq:mode}
\eea
where the integers $k=\hat m n$ labelling these modes are interpreted as  instanton numbers.
As in the analysis of the $w=0$ case the perturbative terms (the $k=0$ terms) arise from two classes of terms:
\begin{itemize}
\item[(i)]  the terms with $n=0$ with a sum over all $\hat m$;
\item[(ii)] the terms with $\hat{m}=0$ with a sum over all $n$.
\end{itemize}
\vskip 0.2cm
{\bf The $n=0$ case}:
\\
In this case we can rewrite the contribution as
\begin{equation}
\cG_{N,0}^{(w)\,(i)} (\tau_2) = \sum_{\hat{m}\in\mathbb{Z}}   \sqrt{\tau_2} \frac{d^{2w}}{d\alpha^{2w}}\Big[ \int_0^\infty \exp\Big( -\frac{(2\hat{m}\sqrt{\pi \tau_2}-i\alpha)^2}{4t} \Big)  t^{w-1/2} B_N(t) dt\Big]_{\alpha=0}\,.
\end{equation}  
{\bf The $\hat m=0$ case}:
\\
In this case we can rewrite the contribution as
\begin{equation}
\cG_{N,0}^{(w)\,(ii)} (\tau_2)  = \sum_{n \in\mathbb{Z}} \sqrt{\tau_2} \frac{d^{2w}}{d\alpha^{2w}}\Big[ \int_0^\infty \exp\Big( -\frac{(2n \sqrt{\pi \tau_2} t+i\alpha)^2}{4t} \Big)  t^{w-1/2} B_N(t) dt\Big]_{\alpha=0}\,.
\label{gniidef}
\end{equation} 
Although it appears that  the $(\hat{m},n)=(0,0)$ term has been double counted, it is fairly simple to show that this actually vanishes thanks to \eqref{intcon}.
 
If we redefine the variable $\alpha$  in \eqref{gniidef} by setting $\alpha=-t\, \tilde \alpha$ so that $d/d\alpha = -t^{-1} d/d\tilde{\alpha}$, and then change variable from $t$ to $1/t$,  \eqref{gniidef} becomes
\begin{equation}
 \cG_{N,0}^{(w)\,(ii)}(\tau_2) = \sum_{n \in\mathbb{Z}}  \sqrt{\tau_2} \frac{d^{2w}}{d{\tilde{\alpha}}^{2w}}\Big[ \int_0^\infty \exp\Big( -\frac{(2n \sqrt{\pi \tau_2}-i \tilde{\alpha})^2}{4 t } \Big)  t^{w-1/2} \frac{B_N\Big(\frac{1}{t}\Big)}{t} dt\Big]_{\tilde \alpha=0}\,,
\end{equation}
which is identical to $ \cG_{N,0}^{(w)\,(i)}(\tau_2)$ using the inversion property $B_N(t) = t^{-1}B_N(t^{-1})$ in \eqref{inverts}.
We conclude that $ \cG_{N,0}^{(w)\,(ii)}(\tau_2)= \cG_{N,0}^{(w)\,(i)} (\tau_2) $, which extends the result previously found when $w=0$.

\subsubsection{Some features of the Yang--Mills perturbation expansion}

Making use of \eqref{modcorr} and the identity in  \eqref{eq:HoloEis}  as well as the definition of the holomorphic Eisenstein series $G_{k}(\tau)$ in  \eqref{holoeisen} it is easy to see that for all $s\leq w$ 
\begin{equation}
\frac{1}{ \tau_2^w} \nabla^w E(s;\tau,\bar\tau) \sim \frac{1}{\tau_2^w} \nabla^{w-s} (\tau_2^{2s} G_{2s}(\tau)) \sim \frac{1}{\tau_2^w} \nabla^{w-s} (\tau_2^{2s}) \sim \tau_2^{s}\,.
\end{equation}
Consequently the $\tau_2^{1-s}$ term in the zero mode of the Eisenstein series with $s\leq w$ does not contribute to the perturbative expansion.
Therefore the first contribution comes from $E(w+1,\tau,\bar\tau)$ so that
\begin{equation}
 \cG_N^{(w)} (\tau,\bar\tau)  \sim O\left(\frac{1}{ \tau_2^w} \nabla^w ( \tau_2^{-w})\right) \sim O(\tau_2^{-w})\,. 
\end{equation}
Therefore we conclude that the perturbation expansion of $\cG_N^{(w)} (\tau,\bar\tau)$ begins at order $\tau_2^{-w}$, i.e. at order $(g_{_{YM}}^2)^w$.

In the $w=0$ case (the four-point correlator) the leading term is of order $\tau_2^0$, which is the free-field contribution  that arose in  \eqref{corrdefs}.   However, following  \cite{Binder:2019jwn} this term cancels out of the supersymmetric localization calculation and the interacting part, which is described by $\cT_N(U,V)$, begins with the one-loop contribution of order $\tau_2^{-1}$.  
If, however, we were to  explicitly integrate the free-field contribution in \eqref{corrdefs} (appropriately normalised) with the measure \eqref{integratedF1}, this would produce a divergent  $\tau_2^0$ coefficient, which can be interpreted as a rational multiple of $\zeta(1)$. This is formally consistent with uniform trascendentality as we will shortly see in \eqref{n2expand}.
 
 On the other hand, for MUV correlators with $w > 0$, we need to include the free-field contributions, which are of order $\tau_2^{-w}$. 
Indeed, as will be explained in section \ref{reloop}, the free part of a $n$-point MUV correlator can also be interpreted as the $(n-4)$-loop correction to the four-point correlator, which provides an efficient method for constructing perturbative loop integrands \cite{Eden:2011we}. 

 Using \eqref{eq:mode} and/or \eqref{gniidef} it is straightforward to determine the perturbative expansion of $\cG_N(\tau,\bar\tau)$  to any order and for any value of $N$.   The following expressions for the perturbative expansion of correlators in the $SU(2)$ theory with different weights (including $w=0, 2, 4$) illustrate the general structure,
 \begin{align}
\cG_{2,0}^{(0)}(\tau_2) &\notag= \cG_{2,0}(\tau_2)=\frac{9 \zeta(3)}{y} -\frac{225 \zeta(5)}{2 y^2} + \frac{2205 \zeta(7)}{2 y^3} - \frac{42525 \zeta(9)}{4 y^4}+O(y^{-5})\,,\\
\cG_{2,0}^{(2)}(\tau_2) & = -\frac{225 \zeta(5)}{ y^2} + \frac{6615 \zeta(7)}{ y^3} - \frac{127575 \zeta(9)}{ y^4}+\frac{8575875 \zeta(11)}{4 y^5}+O(y^{-6})\,,\label{n2expand}\\
\cG_{2,0}^{(4)}(\tau_2) &\notag =-\frac{255150 \zeta(9)}{ y^4} {+}\frac{25727625\zeta(11)}{2 y^5} {-} \frac{1660133475 \zeta(13)}{4 y^6}{+} \frac{22347950625 \zeta(15)}{2 y^7}{+}O(y^{-8})\,,
\end{align}
where $y= \pi \tau_2 $.

It is of interest to exhibit the $N$-dependence of the $SU(N)$  Yang--Mills  perturbation  expansion for generic $N$, which takes the  form
\begin{align}
\label{weak}
\cG_{N, 0}^{(w)}(\tau_2) = &\notag \,\,  (N^2-1)\left[ \frac{3   \,(-1)_w  \zeta (3) a   }{2} -\frac{75 \, (-2)_w\zeta (5) a^2}{8} 
+\frac{735 \,(-3)_w\zeta (7) a^3}{16}  \right.\\
&\notag \left.  -\frac{6615  \,(-4)_w \zeta (9)  \left(1 + \frac{2}{7} N^{-2}\right) a^4 }  {32}   +\frac{114345 \,(-5)_w  \zeta (11) \left(1+ N^{-2} \right) a^5  }{128 }\right. \\
&\notag \left.
 -\frac{3864861 \,(-6)_w\zeta(13) \left(1 +  \frac{25}{11}  N^{-2}+ \frac{4}{11}  N^{-4} \right) a^6}{1024} \right.  \\ 
& \left. +   \frac{32207175 \,(-7)_w\zeta(15) \left(1+ \frac{55}{13} N^{-2}+\frac{332}{143} N^{-4}\right) \,  a^7 }{2048 }+ O(a^{8}) \right] \, ,  
\end{align}
where $a= g_{_{YM}}^2N/ (4 \pi^2) = N/(\pi \tau_2)$ and arbitrary $N\ge 2$. 
Since the Pochhammer symbol $(-n)_w$, with $n\in \mathbb{N}$, vanishes when $n<w$ perturbation theory starts at order $a^{w}\sim \tau_2^{-w}$ for any $N$. As anticipated earlier for $w=0$ the $a^0$ term, which would correspond to a divergent $\zeta(1)$ free-field theory contribution, does not appear.
So we see that although in the case of the $w=0$ correlator, non-planar terms enter the  perturbative expansion  at  four loops,  when $w>0$ non-planar corrections start earlier.  For example, for  $w=2$ the first non-planar correction enters at three loops,
\begin{align}
\label{weakw2}
\cG_{N, 0}^{(2)} (\tau_2)= &\notag \,\,  (N^2-1) \left[ -\frac{75 \,\zeta (5) a^2}{4} 
+\frac{2205 \,\zeta (7) a^3}{8} -\frac{19845  \,\zeta (9)  \left(1 + \frac{2}{7} N^{-2}\right) a^4 }  {8} \right. \\
&\notag \left. +\frac{571725\,  \zeta (11) \left(1+ N^{-2} \right) a^5  }{32 }
 -\frac{57972915 \,\zeta(13) \left(1 +  \frac{25}{11}  N^{-2}+ \frac{4}{11}  N^{-4} \right) a^6}{512} \right.  \\ 
& \left. +   \frac{676350675 \,\zeta(15) \left(1+ \frac{55}{13} N^{-2}+\frac{332}{143} N^{-4}\right) \,  a^7 }{1024 }+ O(a^{8}) \right] \, .
\end{align}
When $w\geq4$ the first non-planar correction enters at leading order (tree-level).\footnote{Note the leading order term arises from Wick contractions of the free theory. }  For example, for $w=4$:
\begin{align}
\label{weak4}
\cG_{N, 0}^{(4)}(\tau_2) = &\notag \,\, (N^2-1)\left[  -\frac{19845  \, \zeta (9)  \left(1 + \frac{2}{7} N^{-2}\right) a^4 }  {4}+\frac{1715175 \, \zeta (11) \left(1+ N^{-2} \right) a^5  }{16 } \right. \\
&\notag \left. 
 -\frac{173918745 \,\zeta(13) \left(1 +  \frac{25}{11}  N^{-2}+ \frac{4}{11}  N^{-4} \right) a^6}{128} \right.  \\ 
& \left. +   \frac{3381753375 \,\zeta(15) \left(1+ \frac{55}{13} N^{-2}+\frac{332}{143} N^{-4}\right) \,  a^7 }{256 }+ O(a^{8}) \right] \, .
\end{align}

\subsection{Instanton and anti-instanton contributions}
\label{sec:instanton}

We will now study the non-zero Fourier modes of  $\cG_{N}^{(w)}(\tau,\bar\tau)$, by using its  expression in terms of non-holomorphic Eisenstein modular forms given in \eqref{modcorr}.    The $k$-instanton contribution to a  single non-holomorphic Eisenstein modular form $E^{(w)}(s;\tau,\bar{\tau})$ (the $k^{th}$ positive Fourier mode with $k>0$), behaves as $e^{2\pi i k \tau} \tau_2^w$ as  $\tau_2\to \infty$, while the $k$ anti-instanton  contribution  (the $k^{th}$ Fourier mode with $k<0$)  vanishes for $s\leq w$ and behaves as  $e^{2\pi i k \tau} \tau_2^{-w}$ when $s > w$.  We will see that the  MUV integrated correlator \eqref{modcorr} has the same general properties.

It follows from  \eqref{bexpand} that the $k^{th}$ Fourier mode for the weight-$w$ integrated correlator can be expressed as
\begin{align}
\cG_{N,k}^{(w)} (\tau, \bar \tau) 
&\label{kmode} = \sum_{ \underset{n\ne 0}{(\hat{m},n)\in\mathbb{Z}^2}}  \sqrt{\tau_2} e^{ 2\pi i \hat{m} n \tau_1} \\
 &\notag\!\!\frac{d^{2w}}{d\alpha^{2w}}\Big[ \int_0^\infty \exp\Big( -\frac{(2n \sqrt{\pi \tau_2} t+i\alpha)^2}{4t}-\frac{(2\hat{m}\sqrt{\pi \tau_2}-i\alpha)^2}{4t}-\frac{\alpha^2}{4t} \Big)  t^{w-1/2}B_N(t) dt \Big]_{\alpha=0} \,.
\end{align}
where $k =\hat{m}n$ and again if we expand the rational function $B_N(t)$ around the origin we find that the integrand can be written as an infinite sum of integrands for $E^{(w)}(s;\tau,\bar\tau)$ in \eqref{eq:mode1}.

Given the definitions of $B_N(t)$ in \eqref{bndef}-\eqref{polydef}, this integral can be explicitly evaluated for fixed mode number $k$ and fixed number of colours $N$.
Alternatively, these non-zero modes of the weight-$w$ integrated correlator can be determined by applying the Cauchy--Riemann derivative   \eqref{eq:CR} to the $w=0$  integrated correlator,  $\cG_N(\tau, \bar \tau)$.  
 
To illustrate the generic features of the instanton terms we will now present some simple explicit examples. These  are the $k=\pm 1$ (charge-one instanton and charge-minus one anti-instanton) sectors of the weight-$2$ and weight-$4$ correlators in the $SU(2)$ theory.  In each case we will present the exact expression together with the first few terms in its perturbative expansion around the $y= \pi \tau_2\to \infty$ limit, 
\ie
\cG_{2,1}^{(2)}(\tau,\bar\tau)   & = e^{2\pi i \tau} \Big[99y^2 -\frac{15}{4} \sqrt{\pi} e^{4y}y^{3/2} (3+56 y) \mbox{erfc}(2\sqrt{y})\Big]  \\
&= e^{2\pi i \tau} \Big(-6y^2+\frac{15}{2}y-\frac{135}{32}+\frac{45}{16 y} +O(y^{-2})\Big)\,,
\label{gex1}
\fe
\ie
\cG_{2,-1}^{(2)} (\tau,\bar\tau) &= e^{ - 2\pi i  \bar  \tau}\Big[ 3 y^2(8y+3)(8y+11) \\
&\qquad\qquad\,- \frac{3}{4}\sqrt{\pi}e^{4y}y^{3/2}(512y^3+960y^2+360y+15) \mbox{erfc}(2\sqrt{y})\Big] \\
& = e^{ - 2\pi i\bar  \tau}\Big(-\frac{135}{256 y^2} +\frac{945}{512 y^3}-\frac{42525}{8192 y^4}+O(y^{-5})\Big)\,,
\label{gex2}
\fe
\ie
\cG_{2,1}^{(4)} (\tau,\bar\tau)& = e^{2\pi i \tau} \Big[\frac{3}{4}y^2(2895+32y(15-4y))-\frac{945}{16}\sqrt{\pi} e^{4y} y^{3/2}(3+88y)\mbox{erfc}(2 \sqrt{y})\Big]   
\\
&= e^{2\pi i \tau} \Big(-96y^4+360 y^3-\frac{855}{2}y^2+\frac{945}{4}y -\frac{14175}{128}+O(y^{-1})\Big)\,,
\label{gex3}
\fe
\ie\label{gex4} 
\cG_{2,-1}^{(4)}(\tau,\bar\tau) &= e^{-2\pi i \bar \tau} \Big[\frac{3 y^2}{4}(2895 + 128 y (165 + 2 y (147 + 8 y (11 + 2 y)))) {-} \frac{3}{16} \sqrt{\pi} e^{4y}y^{3/2} \\
&\!\!\!\!\!\!\!\times (945 + 37800 y + 201600 y^2 + 322560 y^3 + 184320 y^4 + 32768 y^5) \mbox{erfc}(2\sqrt{y})\Big]\,\\
& = e^{-2\pi i \bar \tau}\Big(-\frac{42525}{4096 y^4}+ \frac{1403325}{16384 y^5} -\frac{127702575}{262144 y^6}+O(y^{-7})\Big)\,.
\fe
The general structure of these contributions is in accord with expectations from the analysis of semi-classical instanton contributions to MUV correlators in special cases treated in the literature, see, for example, \cite{Green:1997me, Dorey:1999pd, Green:2002vf}.  These references were all restricted to the holographically related leading low-energy expansion of superstring amplitude, or to leading orders in the $1/N$ expansion of $\cN=4$ SYM correlators, and only considered the semi-classical approximation.  Our  present results go far beyond the semi-classical approximation and apply to any value of $N\ge 2$, but nevertheless some general features are explained by the leading order calculations.  

For example, the fact that the leading power of $g_{_{YM}}^2 \sim \tau_2^{-1}$ in the instanton background is of order $\tau_2^{w}$ is a direct reflection of the presence of 16 superconformal zero modes.   The counting of powers of $\tau_2$ to leading order in $1/\tau_2$ is as follows.  The instanton profile of each operator insertion  involves the product of ($2\Delta-4w$) fermionic zero modes (where $\Delta$ is the dimension of the operator), each contributing a power $\tau_2^{-1/4}$, in addition to the power of $\tau_2$ in the normalisation of each operator.  The leading order instanton contribution to the $n$-point correlator necessarily  absorbs all 16 superconformal  fermion zero modes and is therefore of order $\tau_2^{n-16\times 1/4} = \tau_2^w$ as $\tau_2\to \infty$, as in $\cG_{2,1}^{(2)}$ and $\cG_{2,1}^{(4)}$ exhibited above.  More explicitly, the instanton profile of the operator $\cO_2(x)$ ($\Delta=2$, $w=0$) has four fermionic zero modes, while $\cO_\tau(x)$ ($\Delta=4$, $w=2$) has no fermionic zero modes, and so $\cG_{N, k}^{(w)}(\tau,\bar\tau)$ behaves as
\bea
\langle   \cO_2(x_1, Y_1) \cdots  \cO_2(x_4, Y_4)\, \cO_\tau(x_5)\cdots \cO_\tau(x_{w+4})\rangle  \sim e^{ 2\pi i k \tau} \tau_2^w \, .
\label{ourcorr}
\eea

The contributions to the profiles of operators in an {\it anti}-instanton background acquire more powers of $\tau_2^{-1}$ from two distinct sources. 
\begin{itemize}
\item[(i)]  Firstly, they involve more fermionic modes, which are quasi-zero modes -- these are classical zero modes in the ADHM construction, that arise with non-zero coefficients in the moduli space action when interactions are taken into account (see, for example, \cite{Dorey:1999pd}).  The contributions of such modes  to various correlators is discussed  in \cite{Green:2002vf}. 
\item[(ii)]  Secondly (as is also discussed in detail in \cite{Green:2002vf}),  there are perturbative corrections to the  anti-instanton contribution that arise by Wick contractions of fields inside the operators  in the correlators.  Such contractions appear as propagators joining operators, where the propagator in an instanton background is rather complicated \cite{Green:2002vf}, but has the same power of $\tau_2^{-1}$ as the propagator in a trivial background.  
\end{itemize}
Instead of considering the correlator \eqref{ourcorr} in a $k$-anti-instanton background with $k>0$ we may consider the complex conjugate correlator in a $k$-instanton background with $k>0$, 
\bea
\langle   \cO_2(x_1, Y_1) \cdots  \cO_2(x_4, Y_4)\,\bar \cO_{\bar \tau}(x_5)\cdots\bar \cO_{\bar \tau}(x_{w+4})\rangle \, .
\eea
Since the operator $\bar \cO_{\bar \tau}$ is related to $\cO_\tau$ by the action of eight supercharges, its profile contains the product of eight fermionic zero modes, which may be a mixture of true zero modes and quasi-zero modes.  In general the evaluation of the semi-classical contribution to the correlator involves the sum of both types of contributions, (i) and (ii), described above, together with the contribution of the 16 true superconformal zero modes.   For illustrative purposes we can consider the contribution in which all the fundamental fields in $\bar \cO_{\bar \tau}$  are contracted by propagators in the instanton background and the $16$ true  fermionic zero modes are soaked up by the four $\mathcal{O}_2$ operators. In this contribution the  counting  of the powers of $\tau_2$ has the form
\bea\label{eq:AntiIn}
\langle   \cO_2(x_1, Y_1) \cdots  \cO_2(x_4, Y_4)\,\bar \cO_{\bar \tau}(x_5)\cdots\bar \cO_{\bar \tau}(x_{w+4})\rangle \sim e^{ 2\pi i k \tau} \,\tau_2^{-2w}  \tau_2^w = e^{ 2\pi i k \tau}\tau_2^{-w} \, ,
\eea
where the factor $\tau_2^{-2w}$ arises from $2w$ propagators contracting the fields in $\bar \cO_\tau$'s \footnote{Each $\bar \cO_{\bar \tau}$ contains four fundamental scalar fields, and each propagator contracts two of them, and so $2w$ propagators are needed to contract for $w$ factors of $\bar \cO_\tau$'s.}.  The factor of  $\tau_2^w$ arises from the normalisation factor $\bar O_{\bar \tau} \sim \tau_2$.  

This counting extends to all possible terms involving both integration over quasi-zero modes in the profiles of the operators as in item (i) above, and propagator contractions as in item (ii) above.  All terms contribute the same net power of $\tau_2^{-w}$.\footnote{Although this is not demonstrated explicitly here, closely related examples are given in detail in \cite{Green:2002vf}.}
By taking the complex conjugate of equation \eqref{eq:AntiIn} we then deduce that the correlator \eqref{ourcorr} in a $k$-anti-instanton background, $k>0$, behaves as $ e^{-2\pi i k \bar{\tau}} \tau_2^{-w}$.
We have thus seen that the leading behaviour of $\cG_{N,-k}^{(w)}$ with $k>0$ at large $\tau_2$  is of order  $\tau_2^{-w}$, which in the cases  $\cG_{2,-1}^{(2)}$ and $\cG_{2,-1}^{(4)}$ is in accord with the $\tau_2$-dependence in \eqref{gex2} and \eqref{gex4}.

\section{Large-$N$ expansion} 
\label{largeN}

We will now study large-$N$ expansion of the integrated correlators. We begin by considering the standard 't Hooft limit,  in which $\lambda=g_{_{YM}}^2 N$ is fixed and Yang--Mills instantons are suppressed by factors of $e^{-c N/\lambda}$ for some finite value of $c$. In this limit $\cG_N^{(w)}(\tau,\bar\tau)$  is expressed as a power series in $1/N$, which has the standard interpretation as a genus expansion of the form, 
\bea
\mathcal{G}_N^{(w)}(\tau, \bar \tau) \sim \sum_{g=0}^{\infty} N^{2-2g} \mathcal{G}^{(w, g)} (\lambda)\, , 
\label{Nexpand}
\eea
where $g$ denotes the genus. The expressions for  $\mathcal{G}^{(w, g)} (\lambda)$ and their small-$\lambda$  and large-$\lambda$ expansions are discussed in sections  \ref{smalllam} and \ref{sec:resurg}  for the cases with $g=0,1$.

In order to exhibit the $SL(2,\Z)$ covariance of the correlator it is necessary to consider the large-$N$ limit with fixed $g_{_{YM}}^2$ (sometimes called the ``very strong coupling limit''), in which Yang--Mills instantons play an essential r\^ole.  This  will be the subject of section~\ref{fixedYM}.

\subsection{Large-$N$ and fixed-$\lambda$}

Applying the relation \eqref{muvres} to the perturbative contributions, while ignoring the instanton contribution (which is equivalent to ignoring the $\tau_1$ dependence), the covariant derivative reduces to
\bea 
\label{eq:covar-pert}
\mathcal{D}_{w} = i \left(\tau_2 \frac{\partial}{\partial\tau} -i \frac{w}{2} \right) \rightarrow {1\over 2}\left(\tau_2 \frac{\partial}{\partial \tau_2} + w \right) \,,
\eea
leading to the relation, 
\bea
\mathcal{G}_{N, 0}^{(w)}(\tau_2)  &= &\left(\tau_2 \partial_{\tau_2} + (w-1) \right) \cdots \left(\tau_2 \partial_{\tau_2} + 1 \right)\left(\tau_2 \partial_{\tau_2} \right)\mathcal{G}_{N, 0} (\tau_2) \nn\\
&=& \tau_2 \, \partial_{\tau_2}^{w} \left( \mathcal{G}_{N, 0} (\tau_2) \, \tau_2^{w-1} \right) \, .
\label{redrel}
\eea
  Transforming from $\tau_2$ to the 't Hooft coupling $\lambda= 4\pi N/\tau_2$ and substituting in \eqref{Nexpand} implies, 
\bea\label{eq:diffOpLambda}
\mathcal{G}^{(w,g)}(\lambda) = {1\over \lambda} \, \partial_{ \lambda^{-1}}^{w} \left( \mathcal{G}^{(0, g)} (\lambda)  \lambda^{1-w} \right)\, . 
\eea

We may now consider the series expansions of $\mathcal{G}^{(w,g)}(\lambda)$ at  small $\lambda$ or large $\lambda$.  Note that if we apply the differential operator \eqref{eq:diffOpLambda} to a general function $F(\lambda)$ that has  a small-$\lambda$ expansion of the form $F(\lambda) = \sum_n a_n \lambda^n$ the result is a new Taylor series with coefficients given by
\bea
{1\over \lambda} \, \partial_{ \lambda^{-1}}^{w} \left( F (\lambda)  \lambda^{1-w} \right) = \sum_n a_n {\Gamma(-n+w) \over \Gamma(-n) } \lambda^n \, .
 \label{eq:pertg0}
 \eea
The quantity $\Gamma(-n+w) / \Gamma(-n)$ is the Pochhammer symbol $(-n)_w$ that vanishes when $n<w$, while for $n\geq w$ it can be replaced by the regular expression $(-1)^w (n+1-w)_w$.
Similarly, the  action of the differential operator \eqref{eq:diffOpLambda} on the  large-$\lambda$ expansion, which has the form $F(\lambda) = \sum_n b_n \lambda^{-n-1/2}$, gives rise to an expansion of the form
\bea
{1\over \lambda} \, \partial_{ \lambda^{-1}}^{w} \left( F (\lambda)  \lambda^{1-w} \right)  =  \sum_n b_n {\Gamma(n+1/2+w) \over \Gamma(n+1/2) } \lambda^{-n-1/2} \, .
\eea
So the coefficient for any value of $g$ is determined in terms of the corresponding coefficient in the $w=0$ case.  The factor $\Gamma(n+1/2+w) /\Gamma(n+1/2) = (n+1/2)_w $ is non-vanishing so,  in contrast to the small-$\lambda$ expansion,  the  coefficients in the large-$\lambda$ expansion do not automatically vanish for any value of $n$.

\subsubsection{Small-$\lambda$ expansion and resummation}
\label{smalllam}

We will now apply the above general discussion to concrete examples to obtain explicit results for $\mathcal{G}^{(w, g)} (\lambda)$. To illustrate the structure of these expressions, in the following we will present the results for the first  two genera,  $g=0$ and $g=1$.  
Let us consider the $w=0$ case discussed in \cite{Dorigoni:2021guq} and define $\cG^{(g)}(\lambda) := \cG^{(0,g)}(\lambda)$. 
At leading order in the large-$N$ expansion (i.e. $g=0$),  we have 
\bea
\cG^{(0)} (\lambda) =  \sum_{n=1}^{\infty} \frac{4 (-1)^{n+1} \zeta (2 n+1) \Gamma
   \left(n+\frac{3}{2}\right)^2}{  \pi ^{2n+ 1} \Gamma (n) \Gamma (n+3)}  \lambda ^n\, ,
    \label{small-lam}
\eea
which converges for $|\lambda|<\pi^2$, and can be resummed leading to
\be
\cG^{(0)} (\lambda) =  \lambda \int_0^{\infty} dt \,  t^3 \frac{   _1F_2\left(\frac{5}{2};2,4 \vert-\frac{t^2 \lambda
   }{\pi ^2}\right)}{4 \pi ^2 \, \sinh^2(t)} \, .
   \label{largelam}
\ee
Applying the relation \eqref{eq:pertg0} to \eqref{small-lam}, the resulting series is again convergent for $|\lambda|<\pi^2$, and after again  performing  the resummation, the result is

\bea
\cG^{(w, 0)} (\lambda) = \lambda \int^{\infty}_0 dt \,t^3  \,
   \frac{6(-1)^w\,_1\tilde{F}_2\left(\frac{5}{2};2-w,4\vert-\frac{t^2 \lambda }{\pi
   ^2}\right)}{4\pi^2\sinh^2(t)} \, ,
\label{eq:Gw0alt}
\eea 
where $_1\tilde{F}_2$ is the regularised hypergeometric function, defined by 
\ie
_1\tilde{F}_2(a; b,c \vert z) = {1\over  \Gamma(b)\Gamma(c)} \, _1F_2(a; b,c \vert z) \, ,
\fe 
with $_1F_2$ the usual generalised hypergeometric function.  It is easy to see that $\cG^{(w, 0)} (\lambda)\big|_{w=0}= \cG^{(0)} (\lambda)$.

The $g=1$ contribution to the  $w=0$ correlator considered in \cite{Dorigoni:2021guq} takes the form
\bea
\cG^{(1)} (\lambda) =   \sum_{n=1}^{\infty} \frac{(-1)^n (n-5) (2 n+1) \zeta (2 n+1) \Gamma
   \left(n-\frac{1}{2}\right) \Gamma \left(n+\frac{3}{2}\right)}{24\, \pi ^{2 n+1}  \Gamma (n)^2} \lambda ^n \, ,
   \eea
which converges for $|\lambda| <\pi^2$ and can be resummed to 
\bea
\cG^{(1)} (\lambda){=}-\!\!\lambda\!\! \int_0^{\infty} \!\!dt \,t^3 \frac{   2 \, _2F_3\left(\frac{1}{2},2;1,1,1\vert-\frac{t^2 \lambda }{\pi
   ^2}\right) - 5 \, _1F_2\left(\frac{1}{2};1,2\vert-\frac{t^2
   \lambda }{\pi ^2}\right)-9 J_0\left(\frac{t \sqrt{\lambda }}{\pi }\right){}^2 }{48 \pi ^2 \sinh^2(t) } \label{eq:G1}\, .
   \eea
For non-zero $w$, we find that this result generalises to 
\bea
\cG^{(w,1)} (\lambda)  &= &-\lambda
   \int^{\infty}_0 dt {t^3 \over 48 \pi^2 \sinh^2(t)}\left[ 2 \, _3\tilde{F}_4\left(\frac{1}{2},2,2;2-w,1,1,1\vert-\frac{t^2 \lambda }{\pi
   ^2}\right) \right.\nn\\
   &&\left.-5\,_1\tilde{F}_2\left(\frac{1}{2};1,2-w\vert-\frac{t^2 \lambda }{\pi
   ^2}\right) -9 \, _2\tilde{F}_3\left(\frac{1}{2},2;2-w,1,1\vert-\frac{t^2 \lambda }{\pi
   ^2}\right) 
 \right]\, ,
\label{eq:Gw1alt}
\eea
which reduces to \eqref{eq:G1} in the $w=0$ limit, noticing that $_1\tilde{F}_2\left(\frac{1}{2};1,1\vert-\frac{t^2 \lambda }{\pi
   ^2}\right) = J_0(\frac{t\sqrt{\lambda}}{\pi})^2 $.
Higher-genus terms can be obtained in a similar fashion, and the results have analogous structures to those of $\cG^{(w,0)} (\lambda)$
and $\cG^{(w,1)} (\lambda)$. 

\subsubsection{Large-$\lambda$ expansion and resurgence} 
\label{sec:resurg}
 
In this section, we will consider properties of the integrated MUV correlators in the large-$\lambda$ limit. Using \eqref{largelam} we can straightforwardly obtain the series expansion, for $\cG^{(w,0)} (\lambda)$ and $\cG^{(w,1)} (\lambda)$ from the known results in the $w=0$ case \cite{Dorigoni:2021guq}. Equivalently, one may perform the large-$\lambda$ expansion directly using the integral expressions given in \eqref{eq:Gw0alt} and \eqref{eq:Gw1alt} and the Mellin-Barnes representations of hypergeometric functions. Either way, we find the factorially growing expansion for the $g=0$ coefficient in \eqref{Nexpand},
 \bea
  \cG^{(w,0)} (\lambda)  \sim \frac{\Gamma\Big(w+\frac{1}{2}\Big)}{4\sqrt{\pi}}+  \sum_{n=1}^{\infty}   \frac{   \Gamma
   \left(n-\frac{3}{2}\right) \Gamma
   \left(n+\frac{3}{2}\right) \Gamma (2 n+1) \Gamma\left(n+{1\over 2}+w \right)\zeta (2 n+1) }{2^{2 n-2}  \pi  \,  \Gamma
   (n)^2  \Gamma \left(n+{1\over 2} \right)  \lambda^{n+1/2}}   \,, 
  \label{eq:G0-large}
\eea
 and similarly, for  the $g=1$ term
 \bea
 \cG^{(w,1)} (\lambda)  &\sim & \frac{\Gamma\left(w-{1\over 2} \right) }{32 \sqrt{\pi}}\,\lambda^{1/2}
 \nn\\
 && - \sum_{n=1}^{\infty} \frac{n^2 (2 n+11)    \Gamma
   \left(n+\frac{1}{2}\right) \Gamma \left(n+\frac{3}{2}\right)^2 \Gamma\left(n+{1\over 2}+w \right) \zeta(2n+1)}{24\, \pi ^{\frac{3}{2}} \Gamma (n+2) \Gamma\left(n+{1\over 2}\right)  \lambda^{n+1/2}} \, . 
 \label{sublead}
 \eea
 
 We see that the large-$\lambda$ expansion is asymptotic and not Borel summable for any value of $w$ and  at each order in $1/N$.  We will now see that a  resurgence analysis, following closely the techniques of  \cite{Arutyunov:2016etw, Dorigoni:2021guq}, leads to the non-perturbative completion of these asymptotic series. As in  \cite{Dorigoni:2021guq} this involves understanding the singularities of the asymptotic series after Borel summation. One may also obtain the same result by acting on the $w=0$ expression of \cite{Dorigoni:2021guq}, with appropriate differential operators as discussed earlier.  
 
 We will now consider the convergence properties of the large-$\lambda$ expansion that defines  $\cG^{(w,0)} (\lambda)$  in \eqref{eq:G0-large}. We start by defining a modified Borel transformation \cite{Arutyunov:2016etw}
  \begin{equation}
 \mathcal{B}: \sum_{n=1}^\infty b_n \lambda^{-n-1/2}\rightarrow \sum_{n=1}^\infty \frac{2\pi b_n}{\zeta(2n+1)\Gamma(2n+2)} (2x)^{2n+1} := \hat{\phi}(x)\,,
 \end{equation}
 which, when applied to the asymptotic series \eqref{eq:G0-large}, produces the modified Borel transform
 \begin{equation}
 \hat{\phi}^{(w,0)}(x) = -16\sqrt{\pi}\Gamma\Big( w+\frac{3}{2}\Big) x^3\,_2F_1\Big(-\frac{1}{2},w+\frac{3}{2};1\vert x^2\Big)\,.
 \end{equation}
 Using the key identity 
 \begin{equation}
 \frac{2^{2s-2}}{\Gamma(2s)}\int_0^\infty dx \frac{x^{2s-1}}{\sinh^2(x)} = \zeta(2s-1)\,,
 \end{equation}
 we can then provide an analytic continuation of the formal expansion \eqref{eq:G0-large} in terms of the directional Borel resummation
 \begin{equation}
 \mathcal{S}_\theta \cG^{(w,0)}(\lambda) = \frac{\Gamma\Big(w+\frac{1}{2}\Big)}{4\sqrt{\pi}} +\frac{\sqrt{\lambda}}{\pi}\int_0^{e^{i\theta}\infty} \frac{dx}{4\sinh^2 (x\sqrt{\lambda})} \hat{\phi}^{(w,0)}(x)\,,\label{eq:BorDir}
 \end{equation}
 which defines an analytic function for $\sqrt{\lambda}>0$ when $\theta\in(-\pi/2,\pi/2)$.
 Although \eqref{eq:BorDir} provides an analytic continuation for $\cG^{(w,0)}(\lambda)$ it is neither unique nor is it real for $\sqrt{\lambda}$ positive for any value of integration direction $\theta$.
 This is due to the presence of the branch-cut in the Borel transform $\hat{\phi}^{(w,0)}(x)$ along $[1,\infty]$.
 As anticipated, $\cG^{(w,0)}(\lambda)$ is non-Borel summable and standard resurgence arguments suggest that we are missing exponentially small non-perturbative terms.
 These terms are encoded in the discontinuity of the Borel transform and can be determined in terms of  the so-called Stokes automorphism, which gives
   \begin{equation}
 \lim_{\theta\to 0^+} (\mathcal{S}_{+\theta} - \mathcal{S}_{-\theta})\cG^{(w,0)}(\lambda):= \Delta\cG^{(w,0)}(\lambda) = \frac{\sqrt{\lambda}}{\pi} \int_0^\infty dx \frac{1}{4\sinh^2(x\sqrt{\lambda})} \mbox{Disc}_0\, \hat{\phi}^{(w,0)}(x) \,.
 \label{eq:Stokes}
 \end{equation}
 The discontinuity of the Borel transform can easily be computed for generic $w$ using the known discontinuity for the hypergeometric function arriving at
 \begin{equation}
 \mbox{Disc}_0\, \hat{\phi}^{(w,0)}(x) = \hat{\phi}^{(w,0)}(x+i\,0) - \hat{\phi}^{(w,0)}(x-i\,0) = 16 i \pi \frac{x^3}{(x^2-1)^w}\,_2\tilde{F}_1\Big(\frac{3}{2},-w -\frac{1}{2}; 1-w\vert 1-x^2\Big)\,,
 \label{disc}
 \end{equation}
 with $_2\tilde{F}_1$ again denoting a regularised hypergeometric function.
Following the discussion in \cite{Dorigoni:2021guq}, we can evaluate \eqref{eq:Stokes} using \eqref{disc}, which results in the non-perturbative completion $\Delta\cG^{(w,0)}(\lambda) $. Alternatively, we can apply \eqref{eq:diffOpLambda} to the non-pertubative completion $\Delta\cG^{(0)}(\lambda)$ derived in 
\cite{Dorigoni:2021guq}.  Either method results in  
\bea
\Delta  \cG^{(w,0)} (\lambda) &=&   { i \over \lambda} \partial^w_{\lambda^{-1}} \left[ \lambda^{1-w} \Big( 8 \mbox{Li}_0(e^{-2\sqrt{\lambda}}) + \frac{18\mbox{Li}_1(e^{-2\sqrt{\lambda}})}{\lambda^{1/2}} \right.\nn\\
&&\left. \qquad \qquad + \frac{117 \mbox{Li}_2(e^{-2\sqrt{\lambda}})}{4 \lambda} +\frac{489 \mbox{Li}_3(e^{-2\sqrt{\lambda}})}{16\lambda^{3/2}}+ \ldots \Big) \right]\, .
\eea
It is easy to see that $\Delta \cG^{(w,0)} (\lambda)$ behaves as ${O}(\lambda^{w/2} e^{-2\sqrt{\lambda}} )$. For instance, for  $w=1, 2$, the non-perturbative completions  take the following forms, 
\begin{align}
\Delta \cG^{(1,0)} (\lambda) &=   i\! \left[ 8 \mbox{Li}_{-1}(e^{-2\sqrt{\lambda}}) {\lambda^{1/2}} \!+\! {18\mbox{Li}_0(e^{-2\sqrt{\lambda}})}\!+\! \frac{153 \mbox{Li}_1(e^{-2\sqrt{\lambda}})}{4 \lambda^{1/2}} \!+\!\frac{957 \mbox{Li}_2(e^{-2\sqrt{\lambda}})}{16\lambda}\!+\ldots  \right]\,, \\
\Delta  \cG^{(2,0)} (\lambda) &\notag=   i \!\left[ 8 \mbox{Li}_{{-}2}(e^{-2\sqrt{\lambda}}) {\lambda} \!+\! {22\mbox{Li}_{-1}(e^{-2\sqrt{\lambda}})} {\lambda^{1/2}}  \!+\! \frac{225 \mbox{Li}_0(e^{-2\sqrt{\lambda}})}{4 } \!+\!\frac{1875 \mbox{Li}_1(e^{-2\sqrt{\lambda}})}{16\lambda^{1/2}}\!+ \ldots  \right]\, . 
\end{align}
Similarly, for the $g=1$ coefficient in the large-$N$ expansion, the non-perturbative term is given by
\bea
\! \!  \Delta \cG^{(w,1)} (\lambda) =\!   {-}{ i \over \lambda} \partial^w_{\lambda^{-1}} \!\!\left[ \lambda^{1-w} \Big( {127 \mbox{Li}_0(e^{-2\sqrt{\lambda}}) \over 2^8 } {-} \frac{927\mbox{Li}_1(e^{-2\sqrt{\lambda}})}{2^{12}\lambda^{1/2}} {+} \frac{3897 \mbox{Li}_2(e^{-2\sqrt{\lambda}})}{2^{14} \lambda} \!+ \ldots \Big) \!\right]  ,
\eea
which again behaves as ${O}(\lambda^{w/2} e^{-2\sqrt{\lambda}} )$. Let us also take $w=1, 2$ as examples, 
\ie
\Delta \cG^{(1,1)} (\lambda) &=  - i \left[ {127\mbox{Li}_{-1}(e^{-2\sqrt{\lambda}})  \over 2^8} {\lambda^{1/2}} - \frac{927\mbox{Li}_0(e^{-2\sqrt{\lambda}})}{2^{12}}+ \frac{2043 \mbox{Li}_1(e^{-2\sqrt{\lambda}})}{2^{14} \lambda^{1/2}} + \ldots  \right]\, , \\
\Delta   \cG^{(2,1)} (\lambda) &=   - i \left[ {127\mbox{Li}_{-2}(e^{-2\sqrt{\lambda}})  \over 2^8} {\lambda^{1/2}} - \frac{89\mbox{Li}_{-1}(e^{-2\sqrt{\lambda}})}{2^{12}}- \frac{1665 \mbox{Li}_0(e^{-2\sqrt{\lambda}})}{2^{14} \lambda^{1/2}} + \ldots  \right] \, .
\fe

Using an argument that closely follows appendix D of \cite{Dorigoni:2021guq}, it is easy to prove that the median resummation 
\begin{equation}
\mathcal{S}_{\rm{med}} \cG^{(w,g)} (\lambda) := \lim_{\theta\to 0^+} \Big( \mathcal{S}_{\pm \theta} \cG^{(w,g)}(\lambda) \mp \frac{1}{2}\Delta\cG^{(w,g)}(\lambda)\Big) 
\end{equation}
gives a real expression when $\sqrt{\lambda}>0$, and the analytic continuation is  unambiguous and coincides with the small-$\lambda$ analytic continuation in \eqref{eq:Gw0alt}-\eqref{eq:Gw1alt}.  This demonstrates the importance of the non-perturbative completion $\Delta\cG^{(w,g)}(\lambda)$.

As in \cite{Dorigoni:2021guq}, making use of the AdS/CFT dictionary we can translate these non-perturbative terms into string language, where they should arise from world-sheet instantons. Presumably these would come from a string world-sheet pinned to the $n$ operators in the correlator on the $AdS_5\times S^5$ boundary and stretching into the interior.  However, such a semi-classical picture of these configurations is presently missing.
It is also worth mentioning that similar exponentially suppressed terms have been found \cite{Basso:2007wd} in the large-$\lambda$ expansion of the cusp anomalous dimension in $\mathcal{N}=4$ SYM. 
The strong coupling expansion of this physical quantity, requires a non-perturbative completion with similar, but slightly different, exponentially suppressed terms of order  $\lambda^{1/4} \, e^{-\sqrt{\lambda}/2}$. The cusp anomaly resurgence structure is considerably more complicated than in the present case and was discussed in \cite{Aniceto:2015rua,Dorigoni:2015dha}.  Finally, we notice that  the ``anomalous dimension" associated with the six-point MHV amplitude in $\mathcal{N}=4$ SYM studied in \cite{Basso:2020xts},  behaves as  $e^{-\sqrt{\lambda}}$. It would be of interest to understand the semi-classical origin of the interesting similarities and differences of all these exponentially suppressed terms. 

\subsection{Large-$N$ with fixed-$g_{_{YM}}^2$}
\label{fixedYM}

To study the non-perturbative instanton effects, which are vital for understanding the $SL(2, \mathbb{Z})$ symmetry, we will consider the large-$N$ limit with $g_{_{YM}}^2$ fixed. In this ``very strong limit", when $w=0$, as worked in \cite{Chester:2019jas} and furthered extended in \cite{Dorigoni:2021guq}, the integrated correlator is expanded in terms of non-holomorphic Eisenstein series with half-integer indices, 
\ie
&\label{eq:LargeNHO}\cG_{N}(\tau,\bar\tau) \sim \frac{N^2}{4} - \frac{3N^\half}{2^4}E({\scriptstyle \frac 32}; \tau,\bar\tau)+\frac{45}{2^8 N^\half}E({\scriptstyle \frac 52}; \tau,\bar\tau) \\
& + \frac{3}{N^{\frac{3}{2}}}\Big[\frac{1575}{2^{15}} E({\scriptstyle \frac 72}; \tau,\bar\tau)-\frac{13}{2^{13}}E({\scriptstyle \frac 32}; \tau,\bar\tau) \Big] +\frac{225}{N^{\frac{5}{2}}}\Big[ \frac{441}{2^{18}} E({\scriptstyle \frac 92}; \tau,\bar\tau)  -\frac{5}{2^{16}} E({\scriptstyle \frac 52}; \tau,\bar\tau)\Big]\\
& + \frac{63}{N^{\frac{7}{2}}}\Big[\frac{3898125}{2^{27}}E({\scriptstyle \frac{11}{2}}; \tau,\bar\tau)  -\frac{44625}{2^{25}}E({\scriptstyle \frac 72}; \tau,\bar\tau)+\frac{73}{2^{22}} E({\scriptstyle \frac 32}; \tau,\bar\tau)\Big] \cr
& + \frac{945}{N^{\frac{9}{2}}}\Big[\frac{31216185}{2^{31}}E({\scriptstyle \frac{13}{2}}; \tau,\bar\tau)  -\frac{41895 }{2^{26}}E({\scriptstyle \frac 92}; \tau,\bar\tau)+\frac{1639 }{2^{27}} E({\scriptstyle \frac 52}; \tau,\bar\tau)\Big]  \cr
%& + \frac{33}{N^{\frac{11}{2}}}\Big[\frac{1220198104125}{2^{38}}E({\scriptstyle \frac{15}{2}}; \tau,\bar\tau)  -\frac{12033511875 }{2^{36}}E({\scriptstyle \frac {11}{2} }; \tau,\bar\tau)+\frac{61486425}{2^{34}} E({\scriptstyle \frac 72}; \tau,\bar\tau)- \frac{109447}{2^{32}} E({\scriptstyle \frac 32}; \tau,\bar\tau)\Big] \cr
&+O(N^{-\frac{11}{2}})\, . 
\fe
Using the relations \eqref{zwdef} and \eqref{muvres}, it is straightforward to see that the integrated MUV correlators with the $U(1)_Y$-weight $w$ can be expressed in terms of non-holomorphic Eisenstein series, $ E^{(w)}(s; \tau,\bar\tau)$, 
\begin{align}
&\notag\cG^{(w)}_{N}(\tau,\bar\tau) \sim  - \frac{3 \left(\frac{3}{2}\right)_w N^\half  }{2^4} E^{(w)}({\scriptstyle \frac 32}; \tau,\bar\tau)+\frac{45 \left(\frac{5}{2}\right)_w}{2^8 N^\half}  E^{(w)}({\scriptstyle \frac 52}; \tau,\bar\tau) + \frac{3}{N^{\frac{3}{2}}}\Big[\frac{1575 \left(\frac{7}{2}\right)_w }{2^{15}}   E^{(w)}({\scriptstyle \frac 72}; \tau,\bar\tau)\\
&\label{eq:LargeNHOw} -\frac{13 \left(\frac{3}{2}\right)_w}{2^{13}}   E^{(w)}({\scriptstyle \frac 32}; \tau,\bar\tau) \Big] +\frac{225}{N^{\frac{5}{2}}}\Big[ \frac{441 \left(\frac{9}{2}\right)_w}{2^{18}} E^{(w)}({\scriptstyle \frac 92}; \tau,\bar\tau)  -\frac{5\left(\frac{5}{2}\right)_w}{2^{16}}    E^{(w)}({\scriptstyle \frac 52}; \tau,\bar\tau)\Big]\\
& + \frac{63}{N^{\frac{7}{2}}}\Big[\frac{3898125 \left(\frac{11}{2}\right)_w }{2^{27}}  E^{(w)}({\scriptstyle \frac{11}{2}}; \tau,\bar\tau)  -\frac{44625\left(\frac{7}{2}\right)_w}{2^{25}}    E^{(w)}({\scriptstyle \frac 72}; \tau,\bar\tau)+\frac{73\left(\frac{3}{2}\right)_w}{2^{22}}  E^{(w)}({\scriptstyle \frac 32}; \tau,\bar\tau)\Big] \cr
& + \frac{945}{N^{\frac{9}{2}}}\Big[\frac{31216185 \left(\frac{13}{2}\right)_w }{2^{31}}  E^{(w)}({\scriptstyle \frac{13}{2}}; \tau,\bar\tau)  -\frac{41895\left(\frac{9}{2}\right)_w }{2^{26}}  E^{(w)}({\scriptstyle \frac 92}; \tau,\bar\tau)+\frac{1639 \left(\frac{5}{2}\right)_w}{2^{27}}   E^{(w)}({\scriptstyle \frac 52}; \tau,\bar\tau)\Big]  \notag\\
%& + \frac{33}{N^{\frac{11}{2}}}\Big[\frac{1220198104125}{2^{38}}E({\scriptstyle \frac{15}{2}}; \tau,\bar\tau)  -\frac{12033511875 }{2^{36}}E({\scriptstyle \frac {11}{2} }; \tau,\bar\tau)+\frac{61486425}{2^{34}} E({\scriptstyle \frac 72}; \tau,\bar\tau)- \frac{109447}{2^{32}} E({\scriptstyle \frac 32}; \tau,\bar\tau)\Big] \cr
&+O(N^{-\frac{11}{2}})\, . \nonumber
\end{align}
A few comments are in order. Firstly, the leading large-$N$ term (i.e. the $N^2$ term) in $\cG^{(w)}_{N}(\tau,\bar\tau)$ now disappears due to the action of derivatives. This is consistent with the fact that  the $N^2$ term is associated with the supergravity amplitudes, which cannot violate the $U(1)_Y$ bonus symmetry (or correspondingly the $U(1)$ R-symmetry of type IIB supergravity). Secondly, the results are in accord with the $\alpha'$-expansion of the MUV superamplitudes in type IIB superstring \cite{Green:2019rhz, Green:2020eyj}. In particular,  the non-holomorphic modular Eisenstein series $E^{(w)}({\scriptstyle \frac 32}; \tau,\bar\tau)$ and $E^{(w)}({\scriptstyle \frac 52}; \tau,\bar\tau)$ are associated with the higher-derivative terms $R^4 Z^{w}$ and $d^4R^4 Z^{w}$ (here $Z$ is the dilaton), respectively.  Finally, the result may also be obtained directly from the lattice-sum representation \eqref{lattsum}. This is done by expanding the integrand, especially $B_N(t)$, order by order in $1/N$, as in \cite{Dorigoni:2021guq}. 

\section{Perturbative loop integrands}
\label{reloop}

Apart from the interpretation of \eqref{muvres} in terms of $SL(2,\Z)$-covariant MUV correlators, this equation also leads to a construction of the $w$-loop contribution to the four-point correlator in a manner reminiscent of \cite{Eden:2011we}.  To demonstrate this
we will consider $w$ insertions of $\int \!d^4 x \, \cO_\tau(x)$ in the {\it unintegrated} four-point correlator $ \langle  \cO_2(x_1, Y_1) \cdots \cO_2(x_4, Y_4)\rangle$.  This defines a partially integrated $(4+w)$-point MUV correlator, which has a form given by \eqref{muvres}, but without the integration over $x_1,\dots,x_4$.
The perturbative expansion may be extracted by ignoring the dependence on $\tau_1$.  In other words, by replacing the covariant derivative $\cD_w$  (defined in  \eqref{covdiv}) by ${1\over 2} (\tau_2\, \partial_{\tau_2} + w)$, as given in \eqref{eq:covar-pert}.  With this change   the partially integrated correlator based on \eqref{muvres} reduces to, 
\bea
 && \int d^4 x_{w+4} \cdots d^4 x_{5} \langle  \mathcal{O}_2(x_1, Y_1) \cdots \mathcal{O}_2(x_4, Y_4) \mathcal{O}_{\tau}(x_5)  \cdots \mathcal{O}_{\tau}(x_{w+4})    \rangle = \nn\\
&&  \qquad  \left( \tau_2 \partial_{\tau_2} + (w-1) \right) \cdots  \left( \tau_2 \partial_{\tau_2} + 1 \right)  ( \tau_2 \partial_{\tau_2}) \,  \langle \mathcal{O}_2(x_1, Y_1) \cdots \mathcal{O}_2(x_4, Y_4)  \rangle  \, .
 \label{eq:perturb}
\eea

This equation can be used to determine the perturbative contributions to any MUV correlator, starting  from the lowest order contribution to the four-point correlator.   This follows by considering the perturbative expansion of  
  $\langle \mathcal{O}_2(x_1, Y_1) \cdots \mathcal{O}_2(x_4, Y_4)  \rangle$  in powers of $g_{_{YM}}^2$ (i.e. powers of $\tau_2^{-1}$)
\bea
\langle \mathcal{O}_2(x_1, Y_1) \cdots \mathcal{O}_2(x_4, Y_4)  \rangle = \sum_{L =0 }^{\infty}  \langle \mathcal{O}_2(x_1, Y_1) \cdots \mathcal{O}_2(x_4, Y_4)  \rangle_{L}\, \tau_2^{-L} \, ,
 \label{eq:integrand}
\eea
where $L$ denotes the number of loops in the perturbative expansion, and so the $L$-loop contribution to the correlator is written as  $\langle \mathcal{O}_2(x_1, Y_1) \cdots \mathcal{O}_2(x_4, Y_4)  \rangle_{L}$.

As discussed in section \ref{sec:YMpert}  the product of $SL(2, \mathbb{Z})$ covariant derivatives annihilates the perturbative terms up to order $\tau_2^{-w}$. Indeed, substituting \eqref{eq:integrand} into \eqref{eq:perturb}, leads to 
\bea
&&\sum_{L=w }^{\infty} (-1)^w{\Gamma(L +1) \over \Gamma(L+1-w)} \langle \mathcal{O}_2(x_1, Y_1) \cdots \mathcal{O}_2(x_4, Y_4)  \rangle_{L} \, \tau_2^{-L}  \cr
&=&\, \int d^4 x_{w+4} \cdots d^4 x_{5} \langle  \mathcal{O}_2(x_1, Y_1) \cdots \mathcal{O}_2(x_4, Y_4) \mathcal{O}_{\tau}(x_5)  \cdots \mathcal{O}_{\tau}(x_{w+4})    \rangle  \, .  
\label{eq:integrand2}
\eea
It is easy to check that the lowest-order contribution to the correlator on the right-hand side of this equation is also of order $\tau_2^{-w}$, as follows from Wick contractions of the free theory and the fact that each $\cO_\tau(x)$ is proportional to $\tau_2^{-1}$.  More generally, at $L$ loops in perturbation theory, the correlator on the right-hand side behaves as $\tau_2^{-w-L}$. This is obviously consistent with the left-hand side of the equation.

We can then redefine $L = \ell+w$, with $\ell \geq0$, and by matching the powers of $\tau_2^{-\ell -w}$ on both sides of \eqref{eq:integrand2} we obtain 
\bea 
\label{eq:integrands}
&& \langle \mathcal{O}_2(x_1, Y_1) \cdots \mathcal{O}_2(x_4, Y_4)  \rangle_{\ell+w}\\
&=& \,  (-1)^w{\Gamma(\ell +1) \over \Gamma(\ell + w+1)}   \int d^4 x_{w+4} \cdots d^4 x_{5} \langle  \mathcal{O}_2(x_1, Y_1) \cdots \mathcal{O}_2(x_4, Y_4) \mathcal{O}_{\tau}(x_5)  \cdots \mathcal{O}_{\tau}(x_{w+4})    \rangle_{\ell} \, .  \nn
\eea
The $\ell=0$ case is of particular interest, in that case the right-hand side of this equation is a partially integrated free MUV correlator. The integrand is a rational function of $x_{ij}^2$ with known analytical properties.  It is then identified with the integrand of the $w$-loop contribution to the four-point correlator $\langle \mathcal{O}_2(x_1, Y_1) \cdots \mathcal{O}_2(x_4, Y_4)  \rangle$. The $\ell=0$ combinatorial factor ${1/\Gamma(w+1)}$ accounts for the symmetry factor of the loop integrand.  By further taking out the overall factor $\mathcal{I}_{w+4}$ as in \eqref{ienextra}, the integrand in fact enjoys a $S_{w+4}$ permutation symmetry. All these facts lead to an efficient construction of the correlator to high orders using graph theory \cite{Eden:2011we}.  The idea has been utilised for constructing the loop integrand to a high number of loops \cite{Eden:2011we, Eden:2012tu, Bourjaily:2015bpz, Bourjaily:2016evz, Fleury:2019ydf} in the planar limit and beyond.

An important point is that it is crucial to use the covariant derivative in  \eqref{eq:perturb}, rather than the ordinary derivative $\tau_2 \partial_{\tau_2}$, which was the suggested prescription made in  \cite{Eden:2011we}.   The problem is that  a  product of ordinary derivatives does not   annihilate any low-order terms in the $1/\tau_2$ expansion and therefore $\left( \tau_2 \partial_{\tau_2}  \right)^w \langle \mathcal{O}_2(x_1) \cdots \mathcal{O}_2(x_4)  \rangle$ cannot be identified with the left-hand side of \eqref{eq:perturb}, whose lowest order is $O(\tau_2^{-w})$ as was discussed earlier. In fact, operating with  $\tau_2 \partial_{\tau_2} = {1\over 2} \left[ \left( \tau_2 \partial_{\tau_2} +w \right)  + \left( \tau_2 \partial_{\tau_2} - w \right) \right]$ on the correlator  inserts the sum of  $\int \!d^4x \,\cO_{\tau}(x)$ and $\int \!d^4x \, \cO_{\bar{\tau}}(x)$.  Therefore $\left( \tau_2 \partial_{\tau_2}  \right)^w \langle \mathcal{O}_2(x_1, Y_1) \cdots \mathcal{O}_2(x_4, Y_4)  \rangle$ is related to a linear combination of correlators with insertions of $\mathcal{O}_{\tau}(x)$ and $\mathcal{O}_{\bar{\tau}}(x)$.

\section{Conclusion and Discussion}
\label{discuss} 

 In this paper, we have extended the results of  \cite{Dorigoni:2021bvj, Dorigoni:2021guq}, which concerned properties of an integrated four-point correlator in $\cN=4$ $SU(N)$ SYM, to general $n$-point integrated MUV correlators. In the earlier papers, which were based on the application of supersymmetric localisation techniques \cite{Pestun:2007rz} to the integrated four-point function described in \cite{Binder:2019jwn}, the integrated correlator $\cG_N(\tau,\bar \tau)$  was recast as a two-dimensional lattice sum, which made its modular properties manifest and from which it was simple to analyse its dependence on $N$ and the coupling constant, $\tau$.   Similarly, the $n$-point MUV integrated correlators, $\cG_N^{(n-4)}(\tau,\bar\tau)$ are non-holomorphic modular forms of weight $(n-4, 4-n)$,  which can again be expressed as two-dimensional lattice sums.  They satisfy a Laplace difference equation that is a simple generalisation of the equation satisfied in the $n=4$ case discussed in  \cite{Dorigoni:2021bvj, Dorigoni:2021guq}.  The dependence of $\cG_N^{(n-4)}(\tau,\bar\tau)$  on $N$ and $\tau$ is straightforward to analyse and we presented expansions at weak and strong coupling in the finite-$N$ and large-$N$ limits.  Various systematic features of the dependence of these expansions on the modular weight $w$ were demonstrated in sections~\ref{sec:exact} and \ref{largeN}.  
 
 The results of section  \ref{largeN} reproduce terms in the $1/N$ expansion in the  large-$N$, fixed $g_{_{YM}}^2$ limit  of MUV integrated correlators that were considered in \cite{Green:2019rhz}, and which have a close holographic connection to the leading terms in the low energy expansion of the MUV amplitudes in $AdS_5\times S^5$ in accord with leading terms in the expansion of the flat-space type IIB superstring amplitudes discussed in \cite{Green:2020eyj}.  Integration over the operator positions is crucial for ensuring supersymmetry, which is an essential feature of the localisation arguments.  In this paper, as in the $w=0$ case studied in   \cite{Dorigoni:2021bvj, Dorigoni:2021guq}, our focus was on the integrated correlator defined with the measure in \eqref{integratedF1}.  A second measure \eqref{secondmeasure} that was  introduced in \cite{Chester:2020dja} can be used to  define a different class of integrated $n$-point correlators of the form 
\bea
 \cD_{w-1} \cD_{w-2}\cdots \cD_0\,  \partial_m^4 \log Z_ N (m, \tau, \bar{\tau})  |_{m=0} \,.
\label{newclass}
\eea
The first few  leading terms in the large-$N$ expansion of these correlators, both at fixed $\lambda = g_{_{YM}}^2 N$ and at fixed $g_{_{YM}}^2$, were determined in  \cite{Chester:2020vyz}.  It remains a challenge to formulate such correlators as lattice sums, which would lead to a more thorough elucidation of their properties.

It would also be of interest  to apply these ideas to study exact properties of two-point functions of BPS operators in $\mathcal{N}=2$ supersymmetric theories\footnote{See \cite{Chester:2018aca, Binder:2018yvd, Binder:2019mpb, Agmon:2019imm}  for the application of supersymmetric localisation to the computation of four-point correlation functions in ABJM theory. }at finite coupling and finite $N$, making use of  the methods of supersymmetric localisation  as, for example, in \cite{Baggio:2014ioa, Baggio:2014sna, Gerchkovitz:2014gta, Gerchkovitz:2016gxx, Rodriguez-Gomez:2016ijh, Baggio:2016skg, Billo:2017glv, Bourget:2018fhe, Billo:2019job, Beccaria:2021hvt, Billo:2021rdb}. These two-point correlators are functions of complexified couplings $\tau$ and $\bar {\tau}$, which also transform properly under the modular transformation. Most of the study has been focused on perturbative expansions. 

 Finally, we would like to emphasise the arguments in section~\ref{reloop}, which pointed important differences between partially integrated MUV correlators and the expressions obtained from the Lagrangian insertion method of determining the $\ell$-loop integrand of the four-point correlator of $\cO_2$'s as formulated in \cite{Eden:2011we}.   We find that the $L= (n-4)$-loop correction to  $\langle \cO_2(x_1)\cdots \cO_2(x_4)\rangle$  follows by inserting $(n-4)$ factors of   $\int \!dx_i\,\cO_{\tau}(x_i)$, which corresponds to applying $(n-4)$ {\it covariant} derivatives to the four-point correlator,   (as in \eqref{eq:integrand2}).  By contrast, the procedure advocated in \cite{Eden:2011we}  is to apply $(n-4)$ {\it ordinary} derivatives of the form $g_{_{YM}}^2 \partial/\partial g_{_{YM}}^2$, which  do not contain inhomogeneous terms. This results in the insertion of $(n-4)$ factors of $\int\! d^4x(\cO_\tau (x) + \bar\cO_{\bar\tau}(x))$, which is not of the form  \eqref{eq:integrands}.  However, since in  \cite{Eden:2011we}  the $L$-loop correlator was assumed to be of the form  \eqref{eq:integrands}, their expressions are correct.

\section*{Acknowledgements}

 DD would like to thank the Albert Einstein Institute for the hospitality and support during the writing of this paper.
 MBG has been partially supported by STFC consolidated grant ST/L000385/1.   CW is supported by a Royal Society University Research Fellowship No. UF160350.  
 
 \appendix
%%%%%%%%%%%%%%%%%%%%%%%%%%%%%%

%%%%%%%%%%%%%%%%%%%%%%%%%%%%%%%%
 \section{Brief review of  the integrated four-point correlator}
 \label{review}
 
 In this appendix we will summarise properties of the integrated four-point correlator  \cite{Binder:2019jwn,Chester:2020dja} and in  \cite{Dorigoni:2021bvj, Dorigoni:2021guq}.  
These integrated correlators
 are defined in a manner that preserves certain amount of  supersymmetry and  have the general  form 
 \bea
 \int  \prod_{i=1}^4 dx_i  \, \mu(\{x_i\})\, \langle \cO_2(x_1,Y_1)\,\cO_2(x_2,Y_2)\,\cO_2(x_3,Y_3)\, \cO_2(x_4,Y_4)\rangle\,,
 \label{gndef}
 \eea
  where  $\cO_2(x_i,Y_i)$ is  a superconformal primary in the ${\bf 20'}$ of $SU(4)$ R symmetry.  This is defined by  
  $\cO_2(x, Y) := g_{_{YM}}^{-2} \,\Tr(\varphi^I\varphi^J)\, Y_I Y_J $, 
  where  $\varphi^I$ ($I=1,\dots,6$)  is the scalar field in the $\cN=4$ Yang--Mills multiplet and $Y_I$  is a $SO(6)$  null vector that takes care of the R-symmetry indices.  The dependence on the complex coupling constant
  \bea 
 \tau:=\tau_1+i\tau_2:= \frac{\theta}{2\pi} + \frac{4\pi^2}{g_{_{YM}}^2}\,,
  \label{taudef}
  \eea  
is hidden in the action that enters in the definition of the expectation value  and in the overall normalisation  of $\cO_2(x_i,Y_i)$.  The precise form of the integrated correlator depends on the measure $\mu(\{x_i\})$, which  is defined in such a manner that it preserves supersymmetry. 
 
Following standard conventions the correlator can be expressed in the form  
\bea
\langle \cO_2(x_1, Y_1)\cdots \cO_2(x_4, Y_4)  \rangle = {1\over x_{12}^4 x_{34}^4} \left[\cT_{N,\,\rm free}(U,V;Y_i) + \mathcal{I}_4(U,V; Y_i) \cT_{N}(U,V) \right]  \, ,
\label{corrdefs}
\eea
where $\cT_{N,\, \rm free}$ denotes the free correlator, which can be computed by a simple Wick contraction (see, for example, equation (2.11) of \cite{Eden:2011we}) and will be ignored in the following.     The factor $\cI_4(U,V;Y_i)$ encodes the dependence on the R-symmetry quantum numbers and it is independent of $\tau$ and $N$ (see \cite{Eden:2000bk, Nirschl:2004pa}).  The factor $\cT_N(U,V)$ is the nontrivial part of the correlator.  It is independent of the R-symmetry and is the main consideration in the following.  In these expressions the cross-ratios $U$ and $V$ are defined in the standard manner by
\bea
U = {x_{12}^2 x_{34}^2 \over x_{13}^2 x_{24}^2}\, , \qquad V = {x_{14}^2 x_{23}^2 \over x_{13}^2 x_{24}^2} \, ,
\eea
and $x_{ij}=x_i-x_j$.    The expectation value in  \eqref{corrdefs}
and $\langle \dots \rangle$ is defined by the functional integral 
\bea
\langle \prod_{i=1}^4 {\mathcal{O}}_{i} (x_i,Y_i) \rangle = \int [d \Phi] \, e^{ \int \!d^4x\, \cL(x)} \prod_{r=1}^n {\mathcal{O}}_{i} (x_i,Y_i) \,.
\label{corredef}
\eea
where  $e^{\int \!d^4 x\cL(x)} = e^{-\frac{i}{2\tau_2}\int\! d^4 x (\tau \cO_\tau(x)-\bar\tau \bar\cO_{\bar\tau}(x))}$ and $\cO_\tau(x)$, $\bar \cO_{\bar\tau}(x)$ are the chiral and anti-chiral Lagrangians.

Using the conventions in  \eqref{corrdefs} the first example of an integrated four-point correlator  can be expressed as 
 \bea 
\label{integratedF1}
\cG_{N}(\tau,\bar\tau) : =  - {8\over \pi} \int_0^{\infty} dr \int_0^{\pi} d\theta {r^3 \sin^2(\theta) \over U^2} \cT_{N}(U,V) \, ,
\eea
 where $U = 1+r^2 -2r \cos(\theta)$ and $V=r^2$.  As discussed in \cite{Binder:2019jwn}  this expression arises by considering  \eqref{firstmeasure}, when the R-symmetry charges of the four  operators are chosen  in a manner that sets $\cI_4(U,V; Y_i)=V$. 
 The second example of an integrated correlator of the product of four $\cO_2(x,Y)$'s that preserves supersymmetry was presented in \cite{Chester:2020dja} where it was shown to arise from
 \bea
  \partial_m^4 \log Z_ N (m, \tau, \bar{\tau})  |_{m=0} = - {96\over \pi} \int_0^{\infty} dr \int_0^{\pi} d\theta {r^3 \sin^2(\theta) \over U^2}  \bar{D}_{1111}(U, V) \cT_{N}(U,V) \, ,
  \label{secondmeasure}
  \eea
 instead of \eqref{firstmeasure}. The function $\bar{D}_{1111}(U, V)$ is the so-called $D$-function appears in the context of AdS/CFT duality.  This corresponds to a choice of measure that was described in \cite{Chester:2020dja}.\footnote{Compared to the expression given in equation (2.16) of \cite{Chester:2020dja}, here we have slightly simplified the integration measure using the crossing symmetry of $\bar{D}_{1111}(U, V)$ and $\cT_{N}(U,V)$. } However, in  this paper we will only consider properties of correlators that are based on the first integrated correlator, $\cG_N(\tau,\bar\tau)$,  defined by \eqref{firstmeasure} or, equivalently, by \eqref{integratedF1}.

The partition function of $\cN=2^*$ SYM on $S^4$ was determined in  \cite{Pestun:2007rz} in terms of $SU(N)$ gaussian matrix model integrals over the elements of the Lie algebra $\mathfrak{su}(N)$, which reduce to  $(N-1)$-dimensional integrals over eigenvalues of $SU(N)$ matrices.  The $N$-dependence is therefore encoded in the dimensionality of the integrals, which obscures the analysis of $\cG_N(\tau,\bar\tau)$ for general values of $N$ and $\tau$.
However, the considerations in  \cite{Binder:2019jwn,Chester:2020dja}   emphasised the large-$N$ expansion.  This led to interesting patterns in the properties of the expression  for integrated correlator in the large-$N$ expansion at fixed 't Hooft coupling, where instanton contributions are suppressed exponentially in $N$.  In the large-$N$ limit with fixed $g_{_{YM}}$ considered in \cite{Chester:2019jas,Chester:2020vyz}, Yang--Mills instantons are an important element in ensuring the $SL(2,\Z)$ duality of the correlator.

\subsection*{The lattice sum description of the integrated correlator}
The $N$-dependence of $\cG_N(\tau,\bar\tau)$ was made explicit for all values of $N$ by the reformulation of the integrated correlator in terms of a lattice sum as suggested in  \cite{Dorigoni:2021bvj, Dorigoni:2021guq}
 \bea
\cG_{N} (\tau,\bar\tau)  = {1\over 2}  \sum_{(m,n)\in\mathbb{Z}^2}  \int_0^\infty \exp\Big(- t \pi \frac{|m+n\tau|^2}{\tau_2} \Big) B_N(t) \, dt\,,
\label{gsun}
\eea 
 where $B_N(t)$ has the form
 \bea 
 B_N(t)=\frac{\cQ_N(t)}{(t+1)^{2N+1}}\,,
 \label{bndef}
 \eea
 and where  $\cQ_N(t)$ is a polynomial of degree $2N-1$ that  takes the form
 \begin{align}
&\cQ_N(t)
= -{1\over 2} N (N-1) (1-t)^{N-1} (1+t)^{N+1}  \label{polydef} \\
 &\notag \qquad\left\{ \left(3+  (8N+3t-6) \, t\right ) P_N^{(1,-2)} \left(\frac{1+t^2}{1-t^2}\right)  + \frac{1}  {1+t}   \left(3t^2-8Nt-3 \right) P_N^{(1,-1)}    \left(\frac{1+t^2}{1-t^2}\right)  \right\} \,,
\end{align}
and $P_N^{(\alpha,\beta)} (z)$ is a Jacobi polynomial.   It is significant that the function $B_N(t)$ satisfies the inversion condition 
\bea
B_N(t)= \frac{1}{t} B_N\left(\frac{1}{t}\right)\,,
\label{inverts}
\eea
as well as the integration conditions
\bea
\int_0^\infty B_N(t) \, dt = \frac{N(N-1)}{4} \,,\qquad\quad  \int_0^\infty B_N(t) \, \frac{1}{\sqrt t} \, dt = 0 \,.
\label{intcon}
\eea

  The function $\cG_{N} (\tau,\bar\tau)$ defined in equation \eqref{gsun}  is manifestly invariant under the $SL(2,\Z)$ transformations 
 \begin{equation}
 \tau\to \gamma\cdot \tau = \frac{ a\tau+b}{c\tau +d}\,,\qquad\qquad \gamma = \left(\begin{matrix}a & b\\ c&d\end{matrix}\right) \in SL(2,\Z)\, ,
 \end{equation}
 which is in accord with the expectations of Montonen--Olive duality   \cite{Montonen:1977sn, Witten:1978mh,  Osborn:1979tq}.  
In fact, as shown in \cite{Dorigoni:2021guq}, the expression \eqref{gsun} can be re-expressed as a formal infinite sum of non-holomorphic Eisenstein series with integer indices,
 \begin{align}
\cG_{N}  (\tau,\bar\tau) =  {N(N-1) \over 8}+ {1\over 2} \sum_{s=2}^\infty  c^{(N)}_s E(s; \tau,\bar\tau)\, ,
\label{guborelN-3}
\end{align} 
where the coefficients $c^{(N)}_s$ are defined from $B_{N}(t)$ via the expansion
  \begin{align}
B_{N}(t)  =  \sum_{s=2}^{\infty} c^{(N)}_s {t^{s-1} \over \Gamma(s)} \, .
\label{bexpand}
\end{align}
The definition and properties of non-holomorphic Eisenstein series are reviewed in appendix~\ref{app:Eisendef}.

It was also shown in  \cite{Dorigoni:2021bvj, Dorigoni:2021guq} that the integrated correlator satisfies the Laplace-difference equation 
\begin{equation}
(\Delta_{\tau}-2)\mathcal{G}_N = N^2 ( \mathcal{G}_{N+1}-2 \mathcal{G}_{N}+ \mathcal{G}_{N-1}) -N( \mathcal{G}_{N+1}- \mathcal{G}_{N-1})\,,
\label{lapdiff}
\end{equation}
which connects the integrated correlators for $SU(N)$ gauge group with those of $SU(N+1)$ and $SU(N-1)$.
One consequence is that the expressions for $\cG_N(\tau,\bar\tau)$ with $N>2$ can be determined iteratively in terms of $\cG_2(\tau,\bar\tau)$.

As will be reviewed in appendix~\ref{app:Eisendef}, each non-holomorphic Eisenstein series $E(s; \tau,\bar\tau)$ contains two perturbative zero mode terms, proportional to $\tau_2^{1-s}$ and  $\tau_2^s$, respectively. Using \eqref{guborelN-3}, this leads to power-behaved terms in $1/\tau_2 \sim g_{_{YM}}^2$ and in $\tau_2 \sim 1/g_{_{YM}}^2$ for $\cG_N(\tau,\bar\tau)$.  The series of terms proportional to $\tau_2^{1-s}$ is Borel summable, resulting in a perturbative contribution denoted $\cG^{(i)}_{N, 0}(\tau_2)$,
\begin{equation}
 \cG_{N,0}^{(i)} (\tau_2) =  \sum_{n>0}  \int_0^\infty \exp\Big(- t \pi n^2\tau_2 \Big) \sqrt{\tau_2\,t}\,B_N \Big(\frac{1}{t}\Big) \frac{dt}{t^2} \, .
 \label{eq:G0pG0pp}
\end{equation}
The other series of terms proportional to $\tau_2^s$ and  denoted $ \cG_{N,0}^{(ii)}(\tau_2)$, is evidently ill-defined term by term  in the $g_{_{YM}}^2\to 0$ limit.  However, it has a well-defined Borel sum which is the same expression as \eqref{eq:G0pG0pp}, so that the full perturbative expansion is given by  
\bea
 \cG_{N,0}(\tau_2)= \cG_{N,0}^{(i)} (\tau_2)+ \cG_{N,0}^{(ii)} (\tau_2)= 2\, \cG_{N,0}^{(i)} (\tau_2)\,.
\label{eq:SUN0} 
\eea

The Yang--Mills perturbation theory of $\cG_N(\tau,\bar\tau)$ can be obtained  for any $SU(N)$ group by expanding \eqref{eq:G0pG0pp} in powers of $g_{_{YM}}^2 = 4\pi/\tau_2$.   The results agree with those that have been obtained directly from perturbative $\cN=4$ SYM once their contributions to the correlator are integrated with the appropriate measure.  In fact, explicit results in $\cN=4$ SYM are only available up to three loops \cite{Drummond:2013nda}.  However another feature of exact results in \cite{Dorigoni:2021bvj, Dorigoni:2021guq} is that they demonstrate that non-planar contributions first enter at four loops, which is also a known feature of $\cN=4$ SYM \cite{Eden:2012tu, Fleury:2019ydf}.  It also predicts the pattern of non-planar contributions  at higher loops.  

The large-$N$ expansion of the correlator was also considered in  the 't Hooft limit  in which $\lambda=g_{_{YM}}^2\, N$ is fixed and instantons are suppressed.  The results confirm and extend the results in \cite{Binder:2019jwn}.  In particular they confirm the results of summing the expansion in powers of $1/\lambda$ at large values of $\lambda$.  However, in  \cite{Dorigoni:2021bvj, Dorigoni:2021guq} it was found that  this expansion is not Borel summable and requires a resurgent  completion that is of order $e^{-2\sqrt \lambda}$.  This may be interpreted in the holographic string theory dual as the effects of world-sheet instantons.

The results obtained in  \cite{Binder:2019jwn} concerning the large-$N$ expansion with fixed $g_{_{YM}}^2$ in which instanton contributions play a vital r\^ole, were also extended in \cite{Dorigoni:2021bvj, Dorigoni:2021guq}.   In fact the Laplace-difference equation  \eqref{lapdiff} determines all the terms of higher order in $1/N$ in the large-$N$ expansion once the first two lowest order terms are given.

\section{Non-holomorphic Eisenstein modular forms}
\label{app:Eisendef}

In order to discuss properties of maximal $U(1)$-violating correlators we will here review some features of the particular class of modular forms  that arise in this context, which are extensions of the standard non-holomorphic Eisenstein series, $E(s, \tau,\bar\tau)$. 

\subsection*{Modular covariant derivatives}

The vector space of modular forms, $M_{(w,\hat{w})}$,  with holomorphic and anti-holomorphic weights $(w,\bar{w})$ is defined by
\begin{equation}
f(\tau) \in M_{(w,\hat{w})}\qquad \Longrightarrow \qquad f(\gamma\cdot\tau) = (c\tau+d)^w (c\bar{\tau}+d)^{\hat{w}} f(\tau)\,,
\label{modform}
\end{equation}
for all $\gamma \!=\! \begin{psmallmatrix}a & b\\c & d\end{psmallmatrix} \!\in\! SL(2,\mathbb{Z})$.

Covariant derivatives can be defined to act on this space by changing the modular weights in the following manner,
\bea
\mathcal{D}_w = i \Big(\tau_2 \frac{\partial}{\partial \tau}  - i \frac{w}{ 2}\Big)   \qquad\quad :  M_{(w,\hat{w})}\mapsto M_{(w+1,\hat{w}-1)}\,,
\label{covdiv}
\eea
\bea
\bar{\mathcal{D}}_{\hat{w}} = -i \Big(\tau_2 \frac{\partial}{\partial \bar\tau} + i \frac{\hat{w} }{2}\Big)  \qquad\quad   : M_{(w,\hat{w})}\mapsto M_{(w-1,\hat{w}+1)} \, .
\label{covdivbar}
\eea
Denoting the abstract Chevalley basis for $\mathfrak{sl}(2)$ by $\{\mathcal{D},\bar{\mathcal{D}},H\}$ we have the algebra
\begin{align}
[\mathcal{D}, \bar{\mathcal{D}} ] &= H/2 \,,\\
[\mathcal{D}, H ] & = \mathcal{D}\,,\\
[\bar{\mathcal{D}}, H ] & = -\bar{\mathcal{D}}\,.
\end{align}
The operators $\mathcal{D}\vert_{M_{(w,\hat{w})}} =\mathcal{D}_w$ and $\bar{\mathcal{D}}\vert_{M_{(w,\hat{w})}} =\bar{\mathcal{D}}_{\hat{w}}$ together with $H \vert_{M_{(w,\hat{w})}} = H_{w,\hat{w}} = {1 \over 2}(w - \hat{w})$ and $H_{w,\hat{w}}:M_{(w,\hat{w})}\mapsto M_{(w,\hat{w})}$ form a representation of this algebra.
The Casimir operator is given by
\begin{equation}
\Omega = 2 \mathcal{D} \bar{\mathcal{D}}+ 2 \bar{\mathcal{D}} \mathcal{D}+ H^2 = 4 \mathcal{D}\bar{\mathcal{D}} +H(H-1) = 4 \bar{\mathcal{D}}\mathcal{D} +H(H+1)\,.
\label{cassimirdef}
\end{equation}
More explicitly, restricting to the space $M_{(w,\hat{w})}$,
\begin{align}
\Omega\Big\vert_{M_{(w,\hat{w})}} = \Omega_{w,\hat{w}} &= 4 \mathcal{D}_{w-1}\bar{\mathcal{D}}_{\hat{w}} +\frac{1}{4}(w-\hat{w})(w-\hat{w}-2) \label{casdef1}\\
&= 4 \bar{\mathcal{D}}_{\hat{w}-1}\mathcal{D}_{w} +\frac{1}{4}(w-\hat{w})(w-\hat{w}+2)\,.
\label{casdef2}
\end{align}

We will be interested in the $\hat{w} = -w$ case for which the $SL(2,\Z)$  transformation in \eqref{modform} is a multiplicative  phase.  In this case \eqref{casdef1} and \eqref{casdef2} become
\begin{align}
\Omega\Big\vert_{M_{(w,-w)}} &\label{eq:Lapwmw1}
= 4 \mathcal{D}_{w-1}\bar{\mathcal{D}}_{-w} +w(w-1)\\
&\label{eq:Lapwmw2}
= 4 \bar{\mathcal{D}}_{-w-1}\mathcal{D}_{w} +w(w+1)\,,
\end{align}
will play the r\^ole of Laplacians. In the $w=0$ case these reduce to the standard Laplacian on the hyperbolic plane,\begin{equation}
\Omega\Big\vert_{M_{(0,0)}} = 4 \mathcal{D}_{-1}\bar{\mathcal{D}}_{0}  = -(\tau -\bar{\tau})^2 \partial_\tau \partial_{\bar{\tau}}  = \Delta_{\tau}\, .
\end{equation}

Another useful derivative is the Cauchy-Riemann derivative $\nabla = 2 i \tau_2^2 \partial_\tau$. This acts by changing the modular weights in the following manner,
\bea
\nabla = 2 i \tau_2^2 \frac{\partial}{\partial \tau} : M_{(w,\hat{w})}\mapsto M_{(w, \hat w-2)}\, .
\eea
If  $f\in M_{(0,0)}$ Bol's identity implies
\begin{equation}
\mathcal{D}_{n-1} \mathcal{D}_{n-2} \cdots \mathcal{D}_{0} f = \frac{1}{ (2\tau_2)^n} \nabla^n f\,,
\label{eq:CovCR}
\end{equation}
where both sides are $(n,-n)$ modular forms.  In particular, when $f$ is a non-holomorphic Eisenstein series we have (for $n\in \Z$)
\begin{equation}
\mathcal{D}_{n-1} \mathcal{D}_{n-2} \cdots \mathcal{D}_{0} E(n,\tau,\bar\tau) = \frac{1}{ (2\tau_2)^n} \nabla^n  E(n,\tau,\bar\tau)= \frac{\Gamma(2n)}{\Gamma(n)}  \left(\frac{\tau_2}{2\pi} \right)^n G_{2n}(\tau)\,,
\label{eq:HoloEis}
\end{equation}
with $G_{2n}$ the holomorphic Eisenstein series, which is defined (when $n\ge 2$) by
\begin{equation}
G_{2n}(\tau) = \sum_{p\in\Lambda'} \frac{1}{p^{2n}} = 2\zeta(2n) + \frac{2(2\pi i )^{2n}}{(2n-1)!}\sum_{k>0} \sigma_{2n-1}(k) q^k\,,
\label{holoeisen}
\end{equation}
where $q=e^{2\pi i \tau}$.

More generally, in order to analyse the modes of $E(s;\tau,\bar\tau)$ we will make use of the relation
\begin{equation}
(\pi \nabla)^\ell ( y^k (q^m+\bar{q}^m)) = y^{k+\ell}\Big[(k)_\ell \bar{q}^m + q^m \sum_{s=0}^\ell (-1)^s { \ell \choose s} (k+s)_{\ell-s} (4 m y)^s \Big]\,,\label{eq:CR}
\end{equation}
with $y = \pi \tau_2$.

 \subsubsection*{Laplace eigenvalue equations}
 Non-holomorphic Eisenstein series are solutions of the equation
\bea
\left(\Delta_{\tau} - s\, (s-1) \right) E(s; \tau,\bar\tau) =0\,,
\label{eiseneq}
\eea
where the hyperbolic Laplacian is defined by $\Delta_{\tau} = 4 \tau_2^2(\partial_\tau \, \partial_{\bar \tau})$.  The function $E(s; \tau,\bar\tau)$  is  a $SL(2,\ZZ)$ modular function  that satisfies the asymptotic moderate growth condition  $\lim_{\tau_2\to \infty} E(s; \tau,\bar\tau) < \tau_2^a$, where $a$ is a real number.  The solution  has the form\footnote{We are here following the conventions in \cite{Dorigoni:2021bvj, Dorigoni:2021guq} for the normalisation of $E(s;\tau,\bar\tau)$, in which there is an overall factor of $\pi^{-s}$, which is absent in \cite{Green:2019rhz}.}

\bea
\label{eisendef}
E(s;\tau,\bar\tau) = \sum_{(m,n) \neq\,(0,0)}\frac{\tau_2^s} {\pi^s|m+n\tau|^{2s}}     =   \sum_{k \in\ZZ} \cF_k(s,\tau_2) \, e^{2\pi i k \tau_1}\,,
\label{eisenmodes}
\eea
where the zero Fourier mode consists of the sum of two power behaved terms,
\bea
\cF_0(s,\tau_2)  = \frac{2\zeta(2s) }{\pi^s} \tau_2^s+  \ \frac{2\sqrt \pi \,\Gamma(s-\frac{1}{2}) \zeta(2s-1)}{\pi^s\,\Gamma(s)}\, \tau_2^{1-s} \,,
\label{eisenzero}
   \eea
and the non-zero modes  are D-instanton contributions, which are proportional to $K$-Bessel functions,
\bea
\cF_k(s,\tau_2)  =   \frac{4}{\Gamma(s)}\,  |k|^{s-\half} \, \sigma_{1-2s}(|k|)
\sqrt{\tau_2}\,K(s-\half, 2\pi |k|\tau_2) \,, \  \ \  k \neq 0\,,
\label{nonzeroeisen}
\eea
where the divisor sum is defined by  
\bea \label{eq:divisor-sum}
\sigma_p(k)=\sum_{ {d|k}}   d^{p}  \, , \quad {\rm for} \quad k>0 \, ,
\eea 
and we sum over the positive divisors $d$ of $k$.
We are generally interested in correlators of $n= 4+w$ operators in the stress tensor supermultiplet, which are  proportional to  non-holomorphic Eisenstein modular forms of weight  $(w,-w)$  that are defined by  
\be
\label{dzsn}
\cD_w\, E^{(w)} (s;\tau,\bar\tau)= {s+w\over 2}\, E^{(w+1)} (s;\tau,\bar\tau)\, ,
\label{eisform1}
\ee
and
\be
\label{dbarzsn}
\bar \cD_{-w}\,  E^{(w)} (s;\tau,\bar\tau)= {s-w\over 2}\, E^{(w-1)} (s;\tau,\bar\tau)\, \,,
\label{eisform2}
\ee
using the definition of modular covariant derivatives given in (\ref{covdiv}).  The normalisation factors on the right-hand sides of  \eqref{eisform1} and \eqref{eisform2} are arbitrary, so we have chosen them for later convenience.
Iterating   \eqref{dzsn} leads to the expression 
\be
 E^{(w)} (s;\tau,\bar\tau)= \frac{2^w \Gamma(s) } { \Gamma(s+w)}\, \cD_{w-1} \cdots \cD_0\, E^{(0)} (s;\tau,\bar\tau) = \frac{1}{(s)_w \tau_2^w} \nabla^w E (s;\tau,\bar\tau) \, ,
\label{zwdef}
\ee
where $E^{(0)}(s;\tau,\bar\tau) := E (s;\tau,\bar\tau)$ and $(s)_w= \Gamma(s+w)/\Gamma(s)$ is the Pochhammer symbol. It is straightforward to show that this implies
\bea
E^{(w)} (s;\tau,\bar\tau) = \sum_{(m,n)\neq(0,0)} \Big(\frac{m+n\bar{\tau}}{m+n\tau}\Big)^w \frac{\tau_2^s}{\pi^s|m+n\tau|^{2s}}\,.
 \label{;}
\eea
This expression has arisen previously in the context of the low energy expansion of superstring amplitudes \cite{Green:1997me}  and the large-$N$ expansion of MUV correlators \cite{Green:2019rhz}. 

These modular forms satisfy the Laplace equations
\be
\label{laplaceplusn}
4\bar \cD_{-w-1} \cD_w E_{w} (s,\tau)= (s+w)(s-w-1) \, E_{w}(s,\tau)\, ,
\ee
or, equivalently,
\be
\label{laplaceminusn}
4\cD_{w-1} \bar \cD_{-w} E_{w} (s,\tau)= (s-w)(s+w-1) \, E_{w}(s,\tau)\, .
\ee

 We also note that making use of \eqref{eq:CovCR} and \eqref{eq:HoloEis} when $w\ge s$
$E^{(w)} (s;\tau,\bar\tau)$ can be expressed as
\bea
 E^{(w)} (s;\tau,\bar\tau) &=&  \frac{2^w \Gamma(s) } { \Gamma(s+w)}\, \frac{1}{ (2\tau_2)^w} \nabla^w E(s;\tau,\bar\tau) \nn\\ &=&   \frac{\Gamma(2s) } {\pi^s \Gamma(s+w) }   \tau_2^{-w} \nabla^{w-s}   \, \left(\tau_2^{2s} G_{2s}(\tau) \right)\,,
\label{bolse}
\eea
where $G_{2s} (\tau)$ is a holomorphic Eisenstein series defined in \eqref{holoeisen}.

The following integral representation for  $E^{(w)} (s;\tau,\bar\tau)$ is used in the main text
\cite{Green:1997me}, 
\begin{align}
&2^w \mathcal{D}_{w-1} \mathcal{D}_{w-2} \cdots \mathcal{D}_{0}\, E(s;\tau,\bar\tau)  =(s)_w E^{(w)} (s;\tau,\bar\tau)\nn \\
&= \sum_{(m,n)\neq(0,0)} (s)_w \Big(\frac{m+n\bar{\tau}}{\sqrt{\tau_2/\pi}}\Big)^{2w} \int_0^\infty  e^{ - \frac{\pi t | m +n\tau|^2}{\tau_2}} \frac{t^{s+w-1}}{\Gamma(s+w)} dt   \nn\\
&= \sum_{(m,n)\neq(0,0)} \frac{d^{2w}}{d\alpha^{2w}}\Big[\int_0^\infty  \exp\Big( - \frac{\pi t | m +n\tau|^2}{\tau_2}+  \alpha \frac{\sqrt{\pi}(m+n\bar{\tau})}{\sqrt{\tau_2}}\Big) \frac{t^{s+w-1}}{\Gamma(s)} dt \Big]_{\alpha=0}\,.
\label{intmodeins}
\end{align}

 \subsection{Fourier modes of Eisenstein modular forms}
 \label{modeform}
 The Fourier modes of $E^{(w)} (s;\tau,\bar\tau)$ are defined by
\bea
\label{eisendef}
E^{(w)}(s;\tau,\bar\tau) =\sum_{(m,n)\neq(0,0)} \Big(\frac{m+n\bar{\tau}}{m+n\tau}\Big)^w \frac{\tau_2^s}{\pi^s |m+n\tau|^{2s}}    =   \sum_{k \in\ZZ} \cF^{(w)}_k(s,\tau_2) \, e^{2\pi i k \tau_1}\,,
\label{eisenmodmodes}
\eea
 and can be analysed starting with the representation \eqref{intmodeins}, following a similar procedure to that used in determining the mode coefficients $\cF^{(0)}_k(s,\tau_2)\equiv  \cF_k(s,\tau_2)$ in \eqref{eisenmodes} in the $w=0$ case. This consists of dividing the $(m,n)$ sum into two sectors:

{\bf (i)  $n=0$}.  This gives a contribution to the coefficient of the $\tau_2^{s}$ term in the zero mode, which will be denoted $\cF_0^{(w) (i)}(s,\tau_2)$. When $w\ne 0$ the $(m,n)=(0,0)$ term vanishes and it is useful to Poisson resum the $m$ variable, giving
\bea
\cF_0^{(w) (i)}(s,\tau_2) = \sum_{\hat m\in\mathbb{Z}}   \sqrt{\tau_2} \frac{d^{2w}}{d\alpha^{2w}}\Big[ \int_0^\infty \exp\Big( -\frac{(2\hat m \sqrt{\pi \tau_2}-i\alpha)^2}{4t} \Big)  \frac{t^{s+w-3/2}}{\Gamma(s+w)} dt\Big]_{\alpha=0}\,,
\label{fzeroone}
\eea
where we have set $y=\pi \tau_2$ and $\hat m$ is the variable conjugate to $m$ that enters through the Poisson sum.
After evaluating the $t$  integral and using Riemann's functional equation this gives
\begin{align}
\cF_0^{(w) (i)}(s,\tau_2)  &\notag=\sum_{\hat{m}\in\mathbb{Z}}  \sqrt{\tau_2} \frac{i(-1)^{s+w} 2^{1-2s-2w} \Gamma(\tfrac{1}{2}-s-w)}{\Gamma(s+w)}\frac{d^{2w}}{d\alpha^{2w}} (\alpha+2i \hat{m}\sqrt{\pi \tau_2})^{2s+2w-1} \Big\vert_{\alpha=0}\\
&\notag=\sum_{\hat{m}\in\mathbb{Z}}  \frac{ 8^{-w} (-1)^w \hat{m}^{2s-1} \pi^{s-1/2}\tau_2^s \Gamma(2s+2w)\Gamma(1/2-s-w)}{\Gamma(s+w) \Gamma(2s)}\\
& = \frac{2 \zeta(2s) \tau_2^s}{\pi^{s}}\,.
\label{zerotwo}
\end{align}
The result \eqref{zerotwo} is precisely as expected from the action of the Cauchy-Riemann derivative \eqref{eq:CovCR} on the coefficient of $\tau_2^s$  in the zero mode \eqref{eisenzero} of the Eisenstein series $E(s;\tau,\bar{\tau})$ using \eqref{eq:CR}.

{\bf (ii)  $n\ne 0$}.  After a Poisson summation over $m$ (and changing $n\to -n$) the expression \eqref{intmodeins} becomes
\begin{align}
&\label{eq:mode1}\cF^{(w)}_k (s,\tau_2) = \sum_{ \underset{n\ne 0}{
(\hat{m},n)\in\mathbb{Z}^2}} \sqrt{\tau_2} e^{ 2\pi i \hat{m} n \tau_1}\\
&\notag\qquad\qquad\frac{d^{2w}}{d\alpha^{2w}}\Big[ \int_0^\infty \exp\Big( -\frac{(2n \sqrt{\pi \tau_2} t+i\alpha)^2}{4t}-\frac{(2\hat{m}\sqrt{\pi \tau_2}-i\alpha)^2}{4t}-\frac{\alpha^2}{4t} \Big)  \frac{t^{s+w-3/2}}{\Gamma(s+w)} dt\Big]_{\alpha=0}\,.
\end{align}
 This sector  gives a contribution to the sum over instantons with charges $k= \hat m n$. This  includes the $\hat m=0$ terms that contribute to the coefficient of the $\tau_2^{1-s}$ term in the zero mode, which will be denoted $\cF_0^{(w) (i)}(s,\tau_2)$.  This has the form
\bea
\cF_0^{(w) (ii)}(s,\tau_2)= \sum_{n \neq0}  \sqrt{\tau_2} \frac{d^{2w}}{d\alpha^{2w}}\Big[ \int_0^\infty \exp\Big( -\frac{(2n \sqrt{\pi \tau_2} t+i\alpha)^2}{4t} \Big)  \frac{t^{s+w-3/2}}{\Gamma(s+w)} dt\Big]_{\alpha=0}\,.
\eea
In this case the integral is somewhat more complicated.  After the change of variables $\alpha \to \alpha/ (- i n \sqrt{\pi \tau_2})$ we have
\begin{equation}
\cF_0^{(w) (ii)}(s,\tau_2) = \sum_{n \neq0} \frac{2^{3/2-s-w}(-1)^w n^{1-2s} \pi^{1/2-s} \tau_2^{1-s} }{ \Gamma(s+w)}  \frac{d^{2w}}{d\alpha^{2w}}  \alpha^{s+w-1/2} e^\alpha K_{s+w-1/2}(\alpha) \Big\vert_{\alpha=0}\,,
\label{secondzero}
\end{equation}
The $1/2$-integral K-Bessel function is related to a polynomial in $\alpha$ that takes the form
\begin{equation}
\alpha^{s+w-1/2} e^\alpha K_{s+w-1/2}(\alpha)  = 2^{1/2-s-w}\sqrt{\pi}\sum_{a=0}^{s+w-1} (2\alpha)^{s+w-a-1} \frac{\Gamma(s+w+a)}{a! \Gamma(s+w-a)}\,,
\end{equation} 
and so the $2w$ derivatives evaluated at $\alpha = 0$ simply give
\begin{equation}
\frac{d^{2w}}{d\alpha^{2w}}\Big[ \alpha^{s+w-1/2} e^\alpha K_{s+w-1/2}(\alpha) \Big]_{\alpha=0} = 2^{1/2+w-s} \sqrt{\pi} \frac{\Gamma(2s-1)}{\Gamma(s-w)}\,.
\end{equation}
Substituting in \eqref{secondzero} results in
 \begin{equation}
\cF_0^{(w) (ii)}(s,\tau_2) = \frac{(1-s)_w}{(s)_w} \frac{2 \sqrt{\pi} \Gamma(s-{1 \over 2}) \zeta(2s-1)}{\pi^s \Gamma(s)} \tau_2^{1-s}\,,
\end{equation}
from which we recognise the action of the Cauchy--Riemann  derivative \eqref{eq:CR} on the negative power term in the zero Fourier mode \eqref{eisenzero} of $E (s;\tau,\bar{\tau})$. Noting that  $(1-s)_w= \Gamma(1-s+w)/\Gamma(1-s)=0$ when $w\ge s$ (with $w,s \in {\mathbb N}$), we see that the $\tau_2^{1-s}$ term in $\cF_0^{(w) (ii)}(s,\tau_2)$  is absent when $w \ge s$.

\subsubsection*{Non-zero Fourier modes of Eisenstein modular forms}

We will now consider the non-zero modes in the Fourier expansion \eqref{eisenmodmodes}.  In the $w=0$ case the correlator is a real function and the mode expansion is an expansion in a series of $\cos(2\pi k \tau_1)$ functions, so instantons and anti-instantons contribute equally. The instanton contributions are contained in the $\hat m\ne 0$ terms in the sum in  \eqref{eq:mode1}.
We have to distinguish the contributions of instantons with $ k = \hat{m} n >0$, and anti-instantons with $ k = \hat{m} n <0$.
Performing the integral and making a trivial change of variable $\alpha \to \tilde{\alpha} = i n \sqrt{\pi \tau_2}\alpha$ we arrive at
\begin{align}
\cF^{(w)}_k (s,\tau_2) = &\, \frac{(-1)^w 2^{\frac{5}{2}-s-w} \sigma_{1-2s}(|k|) \pi^{1/2-s}\tau_2^{1-s}}{\Gamma(s)} \nn\\
&   \times\frac{d^{2w}}{d\alpha^{2w}} \left\lbrace \begin{matrix}
e^{+\alpha} (2\pi |k|  \tau_2-\alpha)^{s+w-\frac{1}{2}} K_{s+w-1/2}(2 \pi |k|  \tau_2-\alpha) \Big\vert_{\alpha=0} \,, \qquad k>0\\
e^{-\alpha} (2\pi |k|  \tau_2-\alpha)^{s+w-\frac{1}{2}} K_{s+w-1/2}(2\pi |k|  \tau_2-\alpha)\Big\vert_{\alpha=0} \,, \qquad k<0\,.
\end{matrix}\right.\label{eq:NP}
\end{align}
We note, in particular,  that both for $k>0$ and $k<0$ the Bessel function will produce the expected exponentially suppressed factor $e^{-2 \pi |k| \tau_2}$ which will combine with $e^{2\pi i k \tau_1}$ to produce $q^k$ for $k>0$ and $\bar{q}^{|k|}$ for $k<0$.

Secondly it is once again possible to show that the $\bar{q}$ contribution vanishes identically for all $s\leq w$, while in the case $s=w$ the $q$ and $\bar{q}$ contributions simplify dramatically to 
\begin{equation}\label{eq:wequals}
 \cF^{(s)}_k (s,\tau_2) =\left\lbrace\begin{matrix} \frac{(-1)^s 2^{2s+1}(\pi \tau_2)^{s}}{\Gamma(2s)} \sigma_{2s-1}(k)q^k\,,&& k>0\,,\\
 0\,,&& k<0\,,\end{matrix}\right.
 \end{equation}  as expected from \eqref{eq:HoloEis}.

The expression \eqref{eq:NP} could also have been derived from the action of the Cauchy-Riemann derivative \eqref{eq:CR} on the non-zero Fourier mode of the Eisenstein series.

To illustrate this structure the following is a list of a few  terms with small values of $s$ and $w$ where as usual $y = \pi \tau_2$: 
\begin{align}
\sum_{k\neq0} \cF^{(2)}_k (2,\tau_2) e^{2\pi i k \tau_1} &\nn = \sum_{k>0} \frac{16}{3}k^3 y^2 \sigma_{-3}(k) q^k \,,\\
\sum_{k\neq0} \cF^{(2)}_k (3,\tau_2) e^{2\pi i k \tau_1} &\nn = \sum_{k>0}\left\{ \Big[\frac{4}{3}k^4y^2+\frac{4}{3}k^3y+k^2+\frac{k}{2y}+\frac{1}{8 y^2} \Big]
\sigma_{-5}(k) q^k +\frac{ \sigma_{-5}(k) }{8 y^2}{\bar{q}}^{k}\right\} \,,\\
\sum_{k\neq0} \cF^{(3)}_k (3,\tau_2) e^{2\pi ik \tau_1} & = \sum_{k>0} -\frac{16}{15}k^5 y^3 \sigma_{-5}(k) q^k\,.
\end{align}
These examples illustrate the fact that when $w\ge s$ the anti-instanton contributions are absent, and when $w=s$ the result reduces to \eqref{eq:wequals} (using the fact that $k^{2s-1} \sigma_{1-2s}(k) = \sigma_{2s-1}(k)$). Furthermore, the leading 5instanton contribution is of order $\tau_2^w$ as $\tau_2\to \infty$, while the leading anti-instanton contribution is of order $\tau_2^{-w}$.

\bibliographystyle{ssg}
\bibliography{U1-ref}

\begingroup\raggedright\begin{thebibliography}{10}

\bibitem{Dorigoni:2021bvj}
D.~Dorigoni, M.~B. Green, and C.~Wen, ``{Novel Representation of an Integrated
  Correlator in $\mathcal N$ = 4 Supersymmetric Yang-Mills Theory},'' {\em
  Phys. Rev. Lett.} {\bf 126} (2021), no.~16 161601,
  \href{https://arxiv.org/abs/2102.08305}{{\tt 2102.08305}}.

\bibitem{Dorigoni:2021guq}
D.~Dorigoni, M.~B. Green, and C.~Wen, ``{Exact properties of an integrated
  correlator in $ \mathcal{N} $ = 4 SU(N) SYM},'' {\em JHEP} {\bf 05} (2021)
  089, \href{https://arxiv.org/abs/2102.09537}{{\tt 2102.09537}}.

\bibitem{Binder:2019jwn}
D.~J. Binder, S.~M. Chester, S.~S. Pufu, and Y.~Wang, ``{$ \mathcal{N} $ = 4
  Super-Yang-Mills correlators at strong coupling from string theory and
  localization},'' {\em JHEP} {\bf 12} (2019) 119,
  \href{https://arxiv.org/abs/1902.06263}{{\tt 1902.06263}}.

\bibitem{Pestun:2007rz}
V.~Pestun, ``{Localization of gauge theory on a four-sphere and supersymmetric
  Wilson loops},'' {\em Commun. Math. Phys.} {\bf 313} (2012) 71--129,
  \href{https://arxiv.org/abs/0712.2824}{{\tt 0712.2824}}.

\bibitem{Chester:2019pvm}
S.~M. Chester, ``{Genus-2 holographic correlator on AdS$_{5} \times$ S$^{5}$
  from localization},'' {\em JHEP} {\bf 04} (2020) 193,
  \href{https://arxiv.org/abs/1908.05247}{{\tt 1908.05247}}.

\bibitem{Chester:2019jas}
S.~M. Chester, M.~B. Green, S.~S. Pufu, Y.~Wang, and C.~Wen, ``{Modular
  invariance in superstring theory from $ \mathcal{N} $ = 4
  super-Yang-Mills},'' {\em JHEP} {\bf 11} (2020) 016,
  \href{https://arxiv.org/abs/1912.13365}{{\tt 1912.13365}}.

\bibitem{Montonen:1977sn}
C.~Montonen and D.~I. Olive, ``{Magnetic Monopoles as Gauge Particles?},'' {\em
  Phys. Lett. B} {\bf 72} (1977) 117--120.

\bibitem{Witten:1978mh}
E.~Witten and D.~I. Olive, ``{Supersymmetry Algebras That Include Topological
  Charges},'' {\em Phys. Lett. B} {\bf 78} (1978) 97--101.

\bibitem{Osborn:1979tq}
H.~Osborn, ``{Topological Charges for N=4 Supersymmetric Gauge Theories and
  Monopoles of Spin 1},'' {\em Phys. Lett. B} {\bf 83} (1979) 321--326.

\bibitem{Chester:2020dja}
S.~M. Chester and S.~S. Pufu, ``{Far beyond the planar limit in
  strongly-coupled $ \mathcal{N} $ = 4 SYM},'' {\em JHEP} {\bf 01} (2021) 103,
  \href{https://arxiv.org/abs/2003.08412}{{\tt 2003.08412}}.

\bibitem{Chester:2020vyz}
S.~M. Chester, M.~B. Green, S.~S. Pufu, Y.~Wang, and C.~Wen, ``{New modular
  invariants in $ \mathcal{N} $ = 4 Super-Yang-Mills theory},'' {\em JHEP} {\bf
  04} (2021) 212, \href{https://arxiv.org/abs/2008.02713}{{\tt 2008.02713}}.

\bibitem{Green:1997tv}
M.~B. Green and M.~Gutperle, ``{Effects of D instantons},'' {\em Nucl. Phys. B}
  {\bf 498} (1997) 195--227, \href{https://arxiv.org/abs/hep-th/9701093}{{\tt
  hep-th/9701093}}.

\bibitem{Green:1997as}
M.~B. Green, M.~Gutperle, and P.~Vanhove, ``{One loop in eleven-dimensions},''
  {\em Phys. Lett. B} {\bf 409} (1997) 177--184,
  \href{https://arxiv.org/abs/hep-th/9706175}{{\tt hep-th/9706175}}.

\bibitem{Green:1998by}
M.~B. Green and S.~Sethi, ``{Supersymmetry constraints on type IIB
  supergravity},'' {\em Phys. Rev. D} {\bf 59} (1999) 046006,
  \href{https://arxiv.org/abs/hep-th/9808061}{{\tt hep-th/9808061}}.

\bibitem{Green:1999pu}
M.~B. Green, H.-h. Kwon, and P.~Vanhove, ``{Two loops in eleven-dimensions},''
  {\em Phys. Rev. D} {\bf 61} (2000) 104010,
  \href{https://arxiv.org/abs/hep-th/9910055}{{\tt hep-th/9910055}}.

\bibitem{Green:2005ba}
M.~B. Green and P.~Vanhove, ``{Duality and higher derivative terms in M
  theory},'' {\em JHEP} {\bf 01} (2006) 093,
  \href{https://arxiv.org/abs/hep-th/0510027}{{\tt hep-th/0510027}}.

\bibitem{Intriligator:1998ig}
K.~A. Intriligator, ``{Bonus symmetries of N=4 superYang-Mills correlation
  functions via AdS duality},'' {\em Nucl. Phys. B} {\bf 551} (1999) 575--600,
  \href{https://arxiv.org/abs/hep-th/9811047}{{\tt hep-th/9811047}}.

\bibitem{Green:1999qt}
M.~B. Green, ``{Interconnections between type II superstrings, M theory and N=4
  supersymmetric Yang-Mills},'' {\em Lect. Notes Phys.} {\bf 525} (1999) 22,
  \href{https://arxiv.org/abs/hep-th/9903124}{{\tt hep-th/9903124}}.

\bibitem{Eden:1999gh}
B.~Eden, P.~S. Howe, and P.~C. West, ``{Nilpotent invariants in N=4 SYM},''
  {\em Phys. Lett. B} {\bf 463} (1999) 19--26,
  \href{https://arxiv.org/abs/hep-th/9905085}{{\tt hep-th/9905085}}.

\bibitem{Green:2020eyj}
M.~B. Green and C.~Wen, ``{Maximal U(1)$_{Y}$-violating n-point correlators in
  $ \mathcal{N} $ = 4 super-Yang-Mills theory},'' {\em JHEP} {\bf 02} (2021)
  042, \href{https://arxiv.org/abs/2009.01211}{{\tt 2009.01211}}.

\bibitem{Green:2019rhz}
M.~B. Green and C.~Wen, ``{Modular Forms and $SL(2, {\mathbb Z})$-covariance of
  type IIB superstring theory},'' {\em JHEP} {\bf 06} (2019) 087,
  \href{https://arxiv.org/abs/1904.13394}{{\tt 1904.13394}}.

\bibitem{Howe:1999hz}
P.~S. Howe, C.~Schubert, E.~Sokatchev, and P.~C. West, ``{Explicit construction
  of nilpotent covariants in N=4 SYM},'' {\em Nucl. Phys. B} {\bf 571} (2000)
  71--90, \href{https://arxiv.org/abs/hep-th/9910011}{{\tt hep-th/9910011}}.

\bibitem{Eden:2011we}
B.~Eden, P.~Heslop, G.~P. Korchemsky, and E.~Sokatchev, ``{Hidden symmetry of
  four-point correlation functions and amplitudes in N=4 SYM},'' {\em Nucl.
  Phys. B} {\bf 862} (2012) 193--231,
  \href{https://arxiv.org/abs/1108.3557}{{\tt 1108.3557}}.

\bibitem{Abl:2020dbx}
T.~Abl, P.~Heslop, and A.~E. Lipstein, ``{Towards the Virasoro-Shapiro
  amplitude in AdS$_{5} \times S^{5}$},'' {\em JHEP} {\bf 04} (2021) 237,
  \href{https://arxiv.org/abs/2012.12091}{{\tt 2012.12091}}.

\bibitem{Basu:2004dm}
A.~Basu, M.~B. Green, and S.~Sethi, ``{A Curious truncation of N=4
  Yang-Mills},'' {\em Phys. Rev. Lett.} {\bf 93} (2004) 261601,
  \href{https://arxiv.org/abs/hep-th/0406267}{{\tt hep-th/0406267}}.

\bibitem{Basu:2004nt}
A.~Basu, M.~B. Green, and S.~Sethi, ``{Some systematics of the coupling
  constant dependence of N=4 Yang-Mills},'' {\em JHEP} {\bf 09} (2004) 045,
  \href{https://arxiv.org/abs/hep-th/0406231}{{\tt hep-th/0406231}}.

\bibitem{DiVecchia:2015jaq}
P.~Di~Vecchia, R.~Marotta, M.~Mojaza, and J.~Nohle, ``{New soft theorems for
  the gravity dilaton and the Nambu-Goldstone dilaton at subsubleading
  order},'' {\em Phys. Rev. D} {\bf 93} (2016), no.~8 085015,
  \href{https://arxiv.org/abs/1512.03316}{{\tt 1512.03316}}.

\bibitem{Eden:2012tu}
B.~Eden, P.~Heslop, G.~P. Korchemsky, and E.~Sokatchev, ``{Constructing the
  correlation function of four stress-tensor multiplets and the four-particle
  amplitude in N=4 SYM},'' {\em Nucl. Phys. B} {\bf 862} (2012) 450--503,
  \href{https://arxiv.org/abs/1201.5329}{{\tt 1201.5329}}.

\bibitem{Green:1997me}
M.~B. Green, M.~Gutperle, and H.-h. Kwon, ``{Sixteen fermion and related terms
  in M theory on T**2},'' {\em Phys. Lett. B} {\bf 421} (1998) 149--161,
  \href{https://arxiv.org/abs/hep-th/9710151}{{\tt hep-th/9710151}}.

\bibitem{Dorey:1999pd}
N.~Dorey, T.~J. Hollowood, V.~V. Khoze, M.~P. Mattis, and S.~Vandoren,
  ``{Multi-instanton calculus and the AdS / CFT correspondence in N=4
  superconformal field theory},'' {\em Nucl. Phys. B} {\bf 552} (1999) 88--168,
  \href{https://arxiv.org/abs/hep-th/9901128}{{\tt hep-th/9901128}}.

\bibitem{Green:2002vf}
M.~B. Green and S.~Kovacs, ``{Instanton induced Yang-Mills correlation
  functions at large N and their AdS(5) x S**5 duals},'' {\em JHEP} {\bf 04}
  (2003) 058, \href{https://arxiv.org/abs/hep-th/0212332}{{\tt
  hep-th/0212332}}.

\bibitem{Arutyunov:2016etw}
G.~Arutyunov, D.~Dorigoni, and S.~Savin, ``{Resurgence of the dressing phase
  for AdS5 x S5 },'' {\em JHEP} {\bf 01} (2017) 055,
  \href{https://arxiv.org/abs/1608.03797}{{\tt 1608.03797}}.

\bibitem{Basso:2007wd}
B.~Basso, G.~P. Korchemsky, and J.~Kotanski, ``{Cusp anomalous dimension in
  maximally supersymmetric Yang-Mills theory at strong coupling},'' {\em Phys.
  Rev. Lett.} {\bf 100} (2008) 091601,
  \href{https://arxiv.org/abs/0708.3933}{{\tt 0708.3933}}.

\bibitem{Aniceto:2015rua}
I.~Aniceto, ``{The Resurgence of the Cusp Anomalous Dimension},'' {\em J. Phys.
  A} {\bf 49} (2016) 065403, \href{https://arxiv.org/abs/1506.03388}{{\tt
  1506.03388}}.

\bibitem{Dorigoni:2015dha}
D.~Dorigoni and Y.~Hatsuda, ``{Resurgence of the Cusp Anomalous Dimension},''
  {\em JHEP} {\bf 09} (2015) 138, \href{https://arxiv.org/abs/1506.03763}{{\tt
  1506.03763}}.

\bibitem{Basso:2020xts}
B.~Basso, L.~J. Dixon, and G.~Papathanasiou, ``{Origin of the Six-Gluon
  Amplitude in Planar $N=4$ Supersymmetric Yang-Mills Theory},'' {\em Phys.
  Rev. Lett.} {\bf 124} (2020), no.~16 161603,
  \href{https://arxiv.org/abs/2001.05460}{{\tt 2001.05460}}.

\bibitem{Bourjaily:2015bpz}
J.~L. Bourjaily, P.~Heslop, and V.-V. Tran, ``{Perturbation Theory at Eight
  Loops: Novel Structures and the Breakdown of Manifest Conformality in N=4
  Supersymmetric Yang-Mills Theory},'' {\em Phys. Rev. Lett.} {\bf 116} (2016),
  no.~19 191602, \href{https://arxiv.org/abs/1512.07912}{{\tt 1512.07912}}.

\bibitem{Bourjaily:2016evz}
J.~L. Bourjaily, P.~Heslop, and V.-V. Tran, ``{Amplitudes and Correlators to
  Ten Loops Using Simple, Graphical Bootstraps},'' {\em JHEP} {\bf 11} (2016)
  125, \href{https://arxiv.org/abs/1609.00007}{{\tt 1609.00007}}.

\bibitem{Fleury:2019ydf}
T.~Fleury and R.~Pereira, ``{Non-planar data of $ \mathcal{N} $ = 4 SYM},''
  {\em JHEP} {\bf 03} (2020) 003, \href{https://arxiv.org/abs/1910.09428}{{\tt
  1910.09428}}.

\bibitem{Chester:2018aca}
S.~M. Chester, S.~S. Pufu, and X.~Yin, ``{The M-Theory S-Matrix From ABJM:
  Beyond 11D Supergravity},'' {\em JHEP} {\bf 08} (2018) 115,
  \href{https://arxiv.org/abs/1804.00949}{{\tt 1804.00949}}.

\bibitem{Binder:2018yvd}
D.~J. Binder, S.~M. Chester, and S.~S. Pufu, ``{Absence of $D^4 R^4$ in
  M-Theory From ABJM},'' {\em JHEP} {\bf 04} (2020) 052,
  \href{https://arxiv.org/abs/1808.10554}{{\tt 1808.10554}}.

\bibitem{Binder:2019mpb}
D.~J. Binder, S.~M. Chester, and S.~S. Pufu, ``{AdS$_{4}$/CFT$_{3}$ from weak
  to strong string coupling},'' {\em JHEP} {\bf 01} (2020) 034,
  \href{https://arxiv.org/abs/1906.07195}{{\tt 1906.07195}}.

\bibitem{Agmon:2019imm}
N.~B. Agmon, S.~M. Chester, and S.~S. Pufu, ``{The M-theory Archipelago},''
  {\em JHEP} {\bf 02} (2020) 010, \href{https://arxiv.org/abs/1907.13222}{{\tt
  1907.13222}}.

\bibitem{Baggio:2014ioa}
M.~Baggio, V.~Niarchos, and K.~Papadodimas, ``{tt$^{*}$ equations, localization
  and exact chiral rings in 4d $ \mathcal{N} $ =2 SCFTs},'' {\em JHEP} {\bf 02}
  (2015) 122, \href{https://arxiv.org/abs/1409.4212}{{\tt 1409.4212}}.

\bibitem{Baggio:2014sna}
M.~Baggio, V.~Niarchos, and K.~Papadodimas, ``{Exact correlation functions in
  $SU(2) \mathcal N=2$ superconformal QCD},'' {\em Phys. Rev. Lett.} {\bf 113}
  (2014), no.~25 251601, \href{https://arxiv.org/abs/1409.4217}{{\tt
  1409.4217}}.

\bibitem{Gerchkovitz:2014gta}
E.~Gerchkovitz, J.~Gomis, and Z.~Komargodski, ``{Sphere Partition Functions and
  the Zamolodchikov Metric},'' {\em JHEP} {\bf 11} (2014) 001,
  \href{https://arxiv.org/abs/1405.7271}{{\tt 1405.7271}}.

\bibitem{Gerchkovitz:2016gxx}
E.~Gerchkovitz, J.~Gomis, N.~Ishtiaque, A.~Karasik, Z.~Komargodski, and S.~S.
  Pufu, ``{Correlation Functions of Coulomb Branch Operators},'' {\em JHEP}
  {\bf 01} (2017) 103, \href{https://arxiv.org/abs/1602.05971}{{\tt
  1602.05971}}.

\bibitem{Rodriguez-Gomez:2016ijh}
D.~Rodriguez-Gomez and J.~G. Russo, ``{Large N Correlation Functions in
  Superconformal Field Theories},'' {\em JHEP} {\bf 06} (2016) 109,
  \href{https://arxiv.org/abs/1604.07416}{{\tt 1604.07416}}.

\bibitem{Baggio:2016skg}
M.~Baggio, V.~Niarchos, K.~Papadodimas, and G.~Vos, ``{Large-N correlation
  functions in $ \mathcal{N} $ = 2 superconformal QCD},'' {\em JHEP} {\bf 01}
  (2017) 101, \href{https://arxiv.org/abs/1610.07612}{{\tt 1610.07612}}.

\bibitem{Billo:2017glv}
M.~Billo, F.~Fucito, A.~Lerda, J.~F. Morales, Y.~S. Stanev, and C.~Wen,
  ``{Two-point correlators in $N =2$ gauge theories},'' {\em Nucl. Phys. B}
  {\bf 926} (2018) 427--466, \href{https://arxiv.org/abs/1705.02909}{{\tt
  1705.02909}}.

\bibitem{Bourget:2018fhe}
A.~Bourget, D.~Rodriguez-Gomez, and J.~G. Russo, ``{Universality of Toda
  equation in ${\cal N}=2$ superconformal field theories},'' {\em JHEP} {\bf
  02} (2019) 011, \href{https://arxiv.org/abs/1810.00840}{{\tt 1810.00840}}.

\bibitem{Billo:2019job}
M.~Billo, F.~Fucito, G.~P. Korchemsky, A.~Lerda, and J.~F. Morales,
  ``{Two-point correlators in non-conformal $ \mathcal{N} $ = 2 gauge
  theories},'' {\em JHEP} {\bf 05} (2019) 199,
  \href{https://arxiv.org/abs/1901.09693}{{\tt 1901.09693}}.

\bibitem{Beccaria:2021hvt}
M.~Beccaria, M.~Bill\`o, M.~Frau, A.~Lerda, and A.~Pini, ``{Exact results in a
  $ \mathcal{N} $ = 2 superconformal gauge theory at strong coupling},'' {\em
  JHEP} {\bf 07} (2021) 185, \href{https://arxiv.org/abs/2105.15113}{{\tt
  2105.15113}}.

\bibitem{Billo:2021rdb}
M.~Billo, M.~Frau, F.~Galvagno, A.~Lerda, and A.~Pini, ``{Strong-coupling
  results for $\mathcal{N}=2$ superconformal quivers and holography},''
  \href{https://arxiv.org/abs/2109.00559}{{\tt 2109.00559}}.

\bibitem{Eden:2000bk}
B.~Eden, A.~C. Petkou, C.~Schubert, and E.~Sokatchev, ``{Partial
  nonrenormalization of the stress tensor four point function in N=4 SYM and
  AdS / CFT},'' {\em Nucl. Phys. B} {\bf 607} (2001) 191--212,
  \href{https://arxiv.org/abs/hep-th/0009106}{{\tt hep-th/0009106}}.

\bibitem{Nirschl:2004pa}
M.~Nirschl and H.~Osborn, ``{Superconformal Ward identities and their
  solution},'' {\em Nucl. Phys. B} {\bf 711} (2005) 409--479,
  \href{https://arxiv.org/abs/hep-th/0407060}{{\tt hep-th/0407060}}.

\bibitem{Drummond:2013nda}
J.~Drummond, C.~Duhr, B.~Eden, P.~Heslop, J.~Pennington, and V.~A. Smirnov,
  ``{Leading singularities and off-shell conformal integrals},'' {\em JHEP}
  {\bf 08} (2013) 133, \href{https://arxiv.org/abs/1303.6909}{{\tt 1303.6909}}.

\end{thebibliography}\endgroup
 
\end{document}